\documentclass[structabstract]{aa}  
\usepackage{graphicx}
\usepackage{txfonts}
\usepackage{natbib}

\newcommand{\Mgb}{$Mg\,b$}
\newcommand{\Brie}{$Bri_e$}
\newcommand{\logs}{$\log\sigma$}
\newcommand{\DsB}{$D/B$}
\newcommand{\Hbeta}{$H\beta$}
\newcommand{\NaD}{$NaD$}
\newcommand{\OIII}{$OIII$}
\newcommand{\HmK}{$H$-$K$}
\newcommand{\BmR}{$B$-$R$}
\newcommand{\Kabs}{$K_{abs}$}
\newcommand{\MgbFe}{$[MgbFe]'$}

\newcommand{\MgbsFe}{$Mg\,b/Fe$}
\newcommand{\logre}{$\log{r_e}$}
\newcommand{\Mabs}{$M_{abs}$}
\newcommand{\Mgdeux}{$Mg_2$}
\newcommand{\ldiam}{$\log(diam)$}

\begin{document}

\title{A six-parameter space to describe galaxy diversification}

\subtitle{}

\author{D. Fraix-Burnet \inst{1}
\and T. Chattopadhyay\inst{2}
\and A.K. Chattopadhyay\inst{3}
\and E. Davoust\inst{4}
\and M. Thuillard\inst{5}}

\institute{
Universit\'e Joseph Fourier - Grenoble 1 / CNRS, Institut de Plan\'etologie et d'Astrophysique de Grenoble, BP 53, F-38041 Grenoble cedex 9, France \email{fraix@obs.ujf-grenoble.fr}\\
\and
Department of Applied Mathematics,  Calcutta University, 92 A.P.C. Road, Kolkata  700009, India\\
\and
Department of Statistics, Calcutta University, 35 Ballygunge Circular Road, Kolkata 700019, India\\
\and
Universit\'e de Toulouse, CNRS, Institut de Recherches en Astrophysique et Plan\'etologie, 14 av. E. Belin, F-31400 Toulouse, France\\
\and
La Colline, 2072 St-Blaise, Switzerland\\
}

\date{Received January  2012; accepted June 2012}

\abstract
%context
{
The diversification of galaxies is caused by transforming events such as accretion, interaction, or mergers. These explain the formation and evolution of galaxies, which can now be described by many observables. Multivariate analyses are the obvious tools to tackle the available datasets and understand the differences between different kinds of objects. However, depending on the method used, redundancies, incompatibilities, or subjective choices of the parameters can diminish the usefulness of these analyses. The behaviour of the available parameters should be analysed before any objective reduction in the dimensionality and any subsequent clustering analyses can be undertaken, especially in an evolutionary context.
}%aim
{
We study a sample of 424 early-type galaxies described by 25 parameters, 10 of which are Lick indices, to identify the most discriminant parameters and construct an evolutionary classification of these objects.
}%method0
{
Four independent statistical methods are used to investigate the discriminant properties of the observables and the partitioning of the 424 galaxies: Principal component analysis, K-means cluster analysis, minimum contradiction analysis, and Cladistics.
}%result
{
The methods agree in terms of six parameters: central velocity dispersion, disc-to-bulge ratio, effective surface brightness, metallicity, and the line indices \NaD\ and \OIII. 
The partitioning found using these six parameters, when projected onto the fundamental plane, looks very similar to the partitioning obtained previously for a totally different sample and based only on the parameters of the fundamental plane. Two additional groups are identified here, and we are able to provide some more constraints on the assembly history of galaxies within each group thanks to the larger number of parameters.  We also identify another ``fundamental plane'' with the absolute K magnitude, the linear diameter, and the Lick index \Hbeta. 
We confirm that the \Mgb\ vs velocity dispersion correlation is very probably an evolutionary correlation, in addition to several other scaling relations. 
Finally, combining the results of our two papers, we obtain a classification of galaxies that is based on the transforming processes that are at the origin of the different groups.
}%conclusion 
{
By taking into account that galaxies are evolving complex objects and using appropriate tools, we are able to derive an \emph{explanatory} classification of galaxies, based on the physical causes of the diverse properties of galaxies, as opposed to the  \emph{descriptive} classifications that are quite common in astrophysics.
}


   \maketitle
%
%________________________________________________________________

\section{Introduction}
\label{introduction}

Galaxies are complex and evolving objects. Their diversity appears to increase rapidly with the instrumental improvements that produce huge databases. A good understanding of the physics governing the processes at work within and between the different components of galaxies requires numerical simulations that produce synthetic populations of hopefully realistic objects. The number of physical processes that may operate together with their infinite possible configurations, render the morphological Hubble classification and its equivalents obviously too simple. Morphology, as detailed as it can be determined in the visible, is only one component of the physics of galaxies, and ignores many ingredients of galaxy evolution, such as kinematics and chemical composition \citep[e.g.][]{atlas3d7}. In addition, these classifications do not make full use of the wealth of information that observations and numerical simulations provide.

Multivariate partitioning analyses appear to be the most appropriate tools. One basic tool, the Principal component analysis, is relatively well-known \citep[e.g.][]{cabanac2002, RecioBlanco2006}, but this is not a clustering tool in itself. Many attempts to apply multivariate clustering methods have been made \citep[e.g.][]{Ellis2005,Chattopadhyay2006,Chattopadhyay2007,ChattopadhyayGRB2007, Chattopadhyay2008,Chattopadhyay2009a,Chattopadhyay2009b,FDC09,Almeida2010,Fraix2010}. Sophisticated statistical tools are used in some areas of astrophysics and are being developed steadily, but multivariate analysis and clustering techniques have not yet been widely applied across the community. It is true that the interpretation of the results is not always easy. Some of the reasons are given below.

 Before using the available parameters to derive and compare the physical properties of galaxies, it is important to check whether they can \emph{discriminate} between different kinds of galaxies. A partitioning of objects into robust groups can only be obtained with discriminant parameters. This does not necessarily preclude using other information to help the physical and evolutionary interpretation of the properties of the groups and the relationships between groups. Among the descriptors of galaxies, many come directly from the observations independently of any model. In principle, all the information is contained within the spectrum. However, since it is a huge amount of information, it is usually summarized by broad-band fluxes (magnitudes), slopes (colours), medium-band and line fluxes (e.g. the Lick indices). This dimensionality reduction is often guided by observational constraints or some physical a priori, but not by discriminant (i.e. statistical) properties.

Multivariate partitionings group objects according to global similarities. They yield a descriptive classification of the diversity, but do not provide any explanation of the differences in properties between groups. Modelling or numerical simulations must be used to understand physically the partitioning and the relationships between the groups.

However, galaxy properties are indeed explained by evolution. Mass, metallicity, morphology, colours, etc, are all the result of galaxy evolution. It can thus be expected that the relationships between the groups are driven by evolution. In addition, galaxy (formation and) evolution proceeds through a limited number of transforming processes \citep[monolithic collapse, secular evolution, gravitational interaction, accretion/merger, and sweeping/ejection, see][]{jc1,jc2}. Since they depend on so many parameters (initial conditions, nature of the objects involved, impact parameters, ...), the outcomes of each of them vary a lot, so that diversity is naturally created through evolution of the galaxy populations. This is what we call diversification.

It is easy to see that each of these transforming processes follows a ``transmission with modification'' scheme, because a galaxy is made of stars, gas, and dust, which are both transmitted and modified during the transforming event \citep{jc1,jc2,FCD06}. This is why one might expect a hierarchical organization of galaxy diversity with evolutionary relationships between groups. Cladistics has been shown to be an adequate \citep{jc1,jc2,FCD06} and quite effective \citep[e.g.][]{FDC09,Fraix2010} tool for identifying this hierarchical organization. Instead of a descriptive classification of galaxy diversity, we hope to be able to build an explanatory classification that is physically more informative.

In a previous work \citep{Fraix2010}, we found that the fundamental plane of early-type galaxies is probably generated by diversification. We believe that this result can be of such importance as to deserve dedicated studies to assess its robustness. The present work is novel in several fundamentall ways:
\begin{itemize}
   \item A distinct data set is used, for which more parameters are available, useful for the analyses themselves and also for the subsequent interpretation, once the groups are determined (Sect.~\ref{data}),
   \item  Two additional methods, PCA and MCA, are used (Sect.~\ref{methods}).
   \item A new set of parameters is used for the various partitioning methods  (Sect.~\ref{classification}), which are selected in a rather objective way (Sect.~\ref{parameters}), unlike the ad hoc selection of parameters from longtime conventional wisdom \citep[as followed in][]{Fraix2010}.
   \item  Measurement errors are used in the classification in the case of the cladistic analysis (Sect.~\ref{data} and Appendix~\ref{errors}).
   \item The combination of the partitionings in the present paper and those in \citet{Fraix2010} help us to devise a new scheme for galaxy classification.
\end{itemize}

We present the data in Sect.~\ref{data} before describing the philosophy of our approach with the different methods used to analyse the discriminant properties of the parameters (i.e. their ability to discriminate between different groups) and the partitioning of the sample in Sect.~\ref{methods}. We then give the results of these analyses and the ``winning'' set of parameters (Sect.~\ref{parameters}) that is used for the partitioning  (Sect.~\ref{classification}). We then comment on the discriminant parameters (Sect.~\ref{descriptors}) and detail the group properties  (Sect.~\ref{discussion}). Scaling relations, correlations, and scatter plots are presented in Sect.~\ref{scattercorrel}, and we discuss the well-known fundamental plane of early-type galaxies as well as another fundamental plane, which we discover in this paper, in Sect.~\ref{threedcorr}. Finally, we combine our present result with the one in \citet{Fraix2010} by plotting a cladogram that summarizes the inferred assembly histories of the 
galaxies and thus is a tentative new scheme for classifying galaxies (Sect.~\ref{history}). The conclusion of this study closes this paper (Sect.~\ref{conclusion}).

\section{Data and methods}

\subsection{Data}
\label{data}

We selected 424 fully documented galaxies from the sample of 509 early-type galaxies in the local Universe of \citet{Ogando2008}. As these authors point out, this sample appears to be relatively small compared to those at intermediate redshifts that have been obtained with large surveys (such as the Sloan Digital Sky Survey). However, they do have the advantage of higher quality spectroscopic data and more reliable structural information such as the effective radius. To describe the galaxies, we took from \citet{Ogando2008} the 10 parameters that belong to the set of 25 Lick indices defined by \citet{Worthey1997}: \Hbeta, $Fe5015$, $Mg1$, $Mg_2$, \Mgb, $Fe5270$, $Fe5335$, $Fe5406$, $Fe5709$, and \NaD. From these Lick indices, we computed two other parameters defined as:   $\MgbFe=\sqrt{Mgb*(0.72 * Fe5270 + 0.28 * Fe5335)}$ \citep{Thomas2003} and  $\MgbsFe=Mgb/\left({1\over 2}(Fe5270 + Fe5335)\right)$ \citep{Gonzales1993}, which are indicators of metallicity and light-element abundance, respectively. The 
other parameters taken from \citet{Ogando2008} were the number of companions $n_c$, the morphological type $T$, the line index \OIII, the velocity dispersion (\logs), and the linear effective radius (\logre).

The surface brightness within the effective radius (\Brie) and the disc-to-bulge ratio (\DsB) were taken from \citet{Alonso2003}. The absolute magnitude in B (\Mabs) and the distance of the galaxies were taken from Hyperleda\footnote{http://leda.univ-lyon1.fr/}, which adopts a Hubble constant of 70 km/s/Mpc. The distances to three galaxies not available in Hyperleda were taken from the literature : NGC 1400 (27.7 Mpc, from \citet{Perrett1997}), NGC 4550 (15.49 Mpc from \citet{Mei2007}), and NGC 5206 (3.6Mpc from \citet{Karachentsev2002}). The colour \BmR\ was calculated from the corrected apparent B magnitude in Hyperleda and the total R magnitude given by \citet{Alonso2003}. The linear diameter (\ldiam) was computed from logdc given in Hyperleda. The infrared magnitudes and colours were taken or calculated from  NED\footnote{http://nedwww.ipac.caltech.edu/}.

Altogether, we have 25 parameters to describe the 424 galaxies. However, two parameters were removed from our analyses, namely the number of companions $n_c$ and the morphological type $T$, which are both discrete parameters. More importantly, $n_c$ is not a property of the galaxies, but their local environment, while $T$ is qualitative and subjective. The full set of 25 parameters is naturally used to interpret our results.

We thus used 23 parameters for the analyses in this paper: three are geometrical (\DsB, \logre, and \ldiam), two come from medium-resolution spectra (\logs\ and \OIII), in addition to the ten Lick indices (\Hbeta, $Fe5015$, $Mg1$, $Mg_2$, \Mgb, $Fe5270$, $Fe5335$, $Fe5406$, $Fe5709$, and \NaD) and \MgbFe, \MgbsFe, the six others are broad-band observables (\Brie, the absolute magnitudes in B (\Mabs) and K (\Kabs), the total colours \BmR, $J$-$H$, and \HmK).

\citet{Ogando2008} provides error bars for each of these parameters. However, evaluating the influence of measurement errors on the partitioning is difficult because their multivariate distribution function is unknown. This appears to be a big statistical problem. Fuzzy cluster analyses could perhaps be useful but they are quite complicated to implement, and the very good agreement between all our results indicate that such an investment is unnecessary at this point. In addition, measurement uncertainties can easily be integrated into the cladistic analysis. We thus limit ourselves to two restricted assessments: the influence of two determinations of the distances of galaxies (needed to determine $r_e$) on the result, and the cladistic analysis with errors. These are described in Appendix~\ref{fourparam}. In any case, one should consider that the physical nature of galaxies implies that there are continuous variations in the parameters, hence the partitions are necessarily fuzzy with no rigid boundaries 
between groups. This implies that there is some uncertainty in the placement of the individual objects in the multivariate parameter space.

\subsection{Methods}
\label{methods}

The philosophy of our approach is to use multivariate tools in a first step to select the parameters that can discriminate different groups within the whole sample. These parameters, called discriminant parameters, are then used in a second step to partition the data into several groups. 

In this paper, we use four methods, which are described in more details in Appendix~\ref{appendMeth}. Three of them are used to analyse the parameters: Principal component analysis (PCA, Sect.~\ref{metPCA}), minimum contradiction analysis (MCA, Sect.~\ref{metMCA}), and cladistics (Sect.~\ref{metClad}), while the groupings are performed with the two latter (MCA and cladistics) together with a cluster analysis (CA, Sect.~\ref{metCA}),

The four approaches are all very different in philosophy and technique. Since there is no ideal statistical method, it is useful to compare results obtained with these independent methods. Convergence improves confidence, but since the assumptions behind the different techniques are different, exact agreement cannot be expected. In the end, it is the physics that decides whether a partitioning is informative.

\section{Analyses of the parameters}
\label{parameters}

In this section, we investigate the behaviour of the observables using three of the methods presented above: PCA, MCA, and cladistics. These three multivariate techniques use the parameters directly instead of distance measures, as in the cluster analysis also considered in this paper. Hence, much information can be gained about the parameters themselves, such as their correlations (PCA) or their respective behaviour in the partitioning process (MCA and cladistics). From this information, one can infer the discriminant power of all parameters for the studied sample.

\subsection{Principal component analysis}
\label{parPCA}

\begin{figure}
   \centering
 \includegraphics[width=\columnwidth]{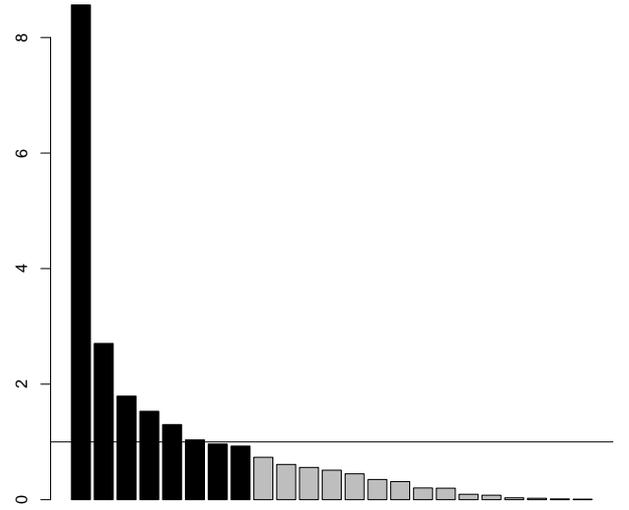}
   \caption{PCA eigenvectors for the sample of 424 objects with 23 parameters. The eigenvalue 1 is indicated by the horizontal line and the eight eigenvectors higher than 1 or so are darkened.} 
    \label{eigenvectors}%
    \end{figure}

We performed a PCA analysis (Sect.~\ref{metPCA}) on the set of 23 parameters. Six principal components (PC or eigenvectors) have eigenvalues greater than 1 and two others are very close to 1 (Fig.~\ref{eigenvectors}), so that eight components describe most (82\%) of the variance of the sample while the first five account for 69\% of the variance.

%%%%%%%%%%%
%%%%%%%%%%%%%

The loadings (i.e. the coefficients of the parameters composing the eigenvectors, Table~\ref{tab_loadings}) give some indications as to which parameters are correlated, redundant, or discriminant. The most important parameters (the first few with the highest loadings in the first eigenvector, and the first parameter for the other eigenvectors) in each PC are:
\begin{enumerate}
\item  \Mgb, \logs, $Mg_2$, \MgbFe, \NaD, $Mg_1$
\item \Brie
\item \OIII
\item \Brie
\item \HmK 
\item $J$-$H$
\item \DsB
\item $Fe5709$
\end{enumerate}

Since the parameters \Mgb, $Mg1$, and $Mg_2$ are so closely related \citep{Burstein1984}, these three quantities are undoubtedly redundant. Moreover, \MgbFe\ depends very much on \Mgb, and is considered a more accurate estimate of the metallicity of a galaxy. Hence, the most important a priori non-redundant parameters are: \logs,  \MgbFe, \NaD, \Brie, \OIII, \HmK, $J$-$H$, \DsB, and $Fe5709$. 

We then performed a second PCA analysis after removing the supposedly redundant parameters (\Mgb, $Mg1$, $Mg_2$, \Mabs) and \logre\ and \ldiam that are affected by the uncertainties in the distance determination (see Appendix~\ref{errors}). We also disregarded $J$-$H$, which has an outlier and otherwise quite constant values. We note that somewhat paradoxically this behaviour could explain why $J$-$H$ appears in the sixth principal component above. We are now left with the 16 parameters \logs, \Brie, \DsB, \Hbeta, $Fe5015$, $Fe5270$, $Fe5335$, $Fe5406$, $Fe5709$, \NaD, \OIII, \HmK, \BmR, \Kabs, \MgbFe, and \MgbsFe. Five eigenvectors have eigenvalues higher than 1 and account for 68\% of the variance. The most important parameters are now:
\begin{enumerate}
\item   \MgbFe, \logs, \NaD
\item $Fe5015$
\item \Brie
\item $Fe5709$
\item \HmK
\end{enumerate}

The agreement is very good with \MgbFe, \logs, \NaD, \Brie, $Fe5709$, and \HmK\ still present, while \OIII\ has disappeared but has loadings very close to those of $Fe5015$ and $Fe5709$ in components 2 and 4, respectively. The disc-to-bulge ratio \DsB\ does not appear to be as important in this analysis. 

\subsection{Minimum contradiction analysis}
\label{parMCA}

The MCA analysis (Sect.~\ref{metMCA}) uses all parameters and explores them to determine the best order (partitioning) that can be obtained. It is possible to derive the discriminant capacity of the parameters according to their respective behaviour as formalised in \citet{TF09}. We find that:

\begin{itemize}
 \item \logs, $Fe5270$, \NaD, \MgbFe, \Brie, \BmR, \OIII, and \DsB\  appear as discriminant parameters;
 \item \logs, \Kabs, \ldiam\ are strongly correlated. 
\end{itemize}

In contrast to PCA, the correlations are not automatically removed, some or all of them may remain. In the present case, the three strongly correlated parameters are not obviously redundant since they are not related by a direct causal relation \citep[see][]{DFB2011}. However, keeping all of them for the MCA analysis does not bring any more discriminating information than keeping only one \citep{TF09}. As a consequence, since \logs\ is listed as one of the discriminant parameters, \Kabs\ and \ldiam\ can be disregarded in the analysis. 

Consequently, the MCA analysis finds eight discriminant parameters.

\subsection{Cladistic analysis}
\label{parClad}

Each cladistic analysis (Sect.~\ref{metClad}) uses and investigates a given set of parameters. To improve our understanding of the 23 parameters, it would be necessary to analyse all possible subsets. Since this require too much computing time, we decided to eliminate obvious redundancies (\Mgb, $Mg1$, $Mg_2$, \Mabs, and \ldiam).  
The index \Hbeta, which is an age indicator for stellar populations older than a few hundred Myr, is problematic for cladistics because age is a property of all groups. This parameter might be able to trace recent transformative events accompanied by starbursts, if it were not for the degeneracy between the age of a younger stellar component and its relative contribution to the total stellar mass or luminosity.  
Guided by the PCA and MCA analyses, we also disregarded $Fe5335$, $Fe5406$, \HmK, \MgbsFe, and $J$-$H$. It is remarkable that \logre\ is not found in the PCA and MCA analyses as a discriminant parameter. We thus also disregarded it here, but kept it for a specific analysis of the fundamental plane together with \logs, \Brie, and $Mg_2$ \citep[see Sect.~\ref{fundplane} and][]{Fraix2010}.

Finally, we studied in more detail the remaining eleven parameters: 
\logs, \DsB, \NaD, \MgbFe, \Brie, \OIII, \Mgb, $Fe5015$, $Fe5270$, $Fe5709$, and \BmR. To find the most discriminant ones in this list, we examined the relative robustness of the trees obtained by cladistic analyses using the eight subsets of these parameters listed in Table~\ref{tab_sets1}. Analyses of each parameter subset were performed with the full sample and several subsamples, and all the results were compared.
The details of our procedure are presented in Sect.~\ref{metClad}. 

\begin{table}
 \centering
% \begin{minipage}{\columnwidth}
  \caption{Subsets of parameters used in cladistic analyses to determine the most discriminant parameters. The names of the subsets include the number of parameters.}
     \label{tab_sets1}
\begin{tabular}{ll}
      \hline
Subset &  Parameters  \\
                  \hline
4cA  & \logs\ \DsB\ \NaD\ \MgbFe  \\
5c    & \logs\ \DsB\ \NaD\ \MgbFe\ \Brie  \\
5cA  & \logs\ \DsB\ \NaD\ \MgbFe\ \Mgb \\
6c    & \logs\ \DsB\ \NaD\ \MgbFe\ \Brie\ \OIII \\
6cA  & \logs\ \DsB\ \NaD\ \MgbFe\ \Mgb\ \Brie  \\
7c    & \logs\ \DsB\ \NaD\ \MgbFe\ \Brie\ \OIII\ $Fe5015$ \\
8c    & \logs\ \DsB\ \NaD\ \MgbFe\ \Brie\ \OIII\ $Fe5015$\ \Mgb\\
10c  &\logs\ \DsB\ \NaD\ \MgbFe\ \Brie\ \OIII\ \Mgb\ $Fe5270$ \\
        & \hfill $Fe5709$\ \BmR  \\
                 \hline
\end{tabular}
%\end{minipage}
\end{table}

Five or six discriminant parameters are favored by the cladistic analysis because the results are then more stable.
The trees from subsets 5c and 6c are in very good agreement, that of 6c being almost entirely structured, and the result for subset 6c is in very good agreement with the cluster analysis (Sect.~\ref{classification}). 

We conclude that the most discriminant 
set of parameters from the cladistic analyses is that of 6c, namely \logs, \DsB, \NaD, \MgbFe, \Brie, and \OIII.

\subsection{Final set of discriminant parameters}
\label{parconcl}

The PCA, MCA, and cladistic analyses agree in terms of the five parameters \logs,  \MgbFe, \NaD, \Brie, and \OIII, while they globally identify five to eight discriminant parameters. The cladistic and MCA analyses find that \DsB\ is an important parameter, which appears only weakly in PCA. \BmR\ is discriminant in MCA only, while the iron indices appear with $Fe5015$ and $Fe5709$ on one side (PCA), and $Fe5270$ on the other (MCA). None of these four parameters are preferred in the cladistic analyses. 

Hence, we select the consensual six parameters \logs, \DsB, \NaD, \MgbFe, \Brie, and \OIII\ for partitioning analyses of our sample.

\section{Partitioning the sample galaxies}
\label{classification}

%%%%%%%%%%%
   \begin{figure}
   \centering
 \includegraphics[width=\columnwidth]{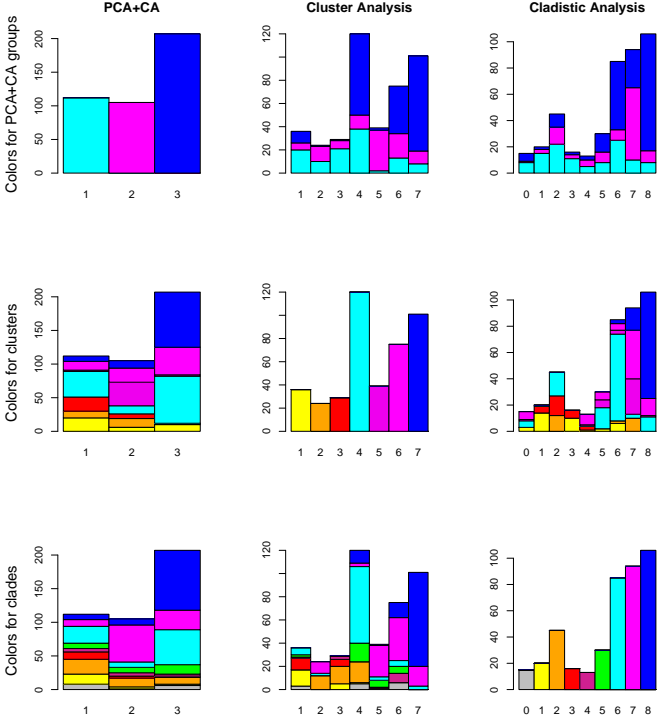}
   \caption{Comparison of analyses with three different methods: PCA+CA (left panels), cluster analysis (middle panels), and cladistics (right panels). The MCA result is not shown since it can be easily compared to the cladistic result (see Sect.~\ref{classcomp}). In the first row, the colours identify the eight groups found in the cladistic analysis; in the second row, they identify the seven groups of the cluster analysis and, in the third row, the three groups found in PCA+CA.
 The colours for the cluster and PCA+CA groups are chosen to more easily visualize the agreement with the cladistic partitioning.} 
    \label{figcomppart}%
    \end{figure}
%%%%%%%%%%%%%

We now compare the partitioning obtained with four methods: a cluster analysis using eight principal components (Sect.~\ref{classPCAClus}), a cluster analysis (Sect.~\ref{classClus}), a MCA optimisation (Sect.~\ref{classMCA}), and a cladistic analysis (Sect.~\ref{classClad}), the latter three using the six parameters listed in Sect.~\ref{parconcl}. The partitionings are compared at the end of this section and in Fig.~\ref{figcomppart}. The order of the groups for cladistics is essentially dictated by the tree (and its rooting, see Sect.~\ref{classClad}), while for the other methods, the order was arbitrarily chosen to correspond as much as possible to the cladistic order.

\subsection{PCA plus cluster analysis}
\label{classPCAClus}

A cluster analysis (Sect.~\ref{metCA}) was performed using the eight PCs obtained by PCA (Sect.~\ref{parPCA} and Sect.~\ref{metPCA}). In this paper, this analysis is denoted PCA+CA. Three groups are found and labelled PCACA1, PCACA2, and PCACA3. The first two contain about 100 objects, while the third is about twice as big (Fig.~\ref{figcomppart}).

As noted in Sect.~\ref{metPCA}, the use of principal components in multivariate clustering very likely obscures a significant part of the underlying physics since it suppresses all correlations, even those that are due to hidden parameters or independent evolutions (see Sect.~\ref{discussion}). We present this result here mainly as an illustration of this point.

\subsection{Cluster analysis}
\label{classClus}

A cluster analysis (Sect.~\ref{metCA}) was performed with the six parameters listed in Sect.~\ref{parconcl}. Seven groups were found and named Clus1 to Clus7. There are three large groups, with about 80 to 120 objects. The other contain from 20 to 40 objects (Fig.~\ref{figcomppart}).

\subsection{Minimum contradiction analysis}
\label{classMCA}

With the six parameters listed in Sect.~\ref{parconcl}, the MCA analysis performs an optimisation of the order to minimise the contradiction (Sect.~\ref{metMCA}). The result is four groups, and maybe two others. The groups are globally very fuzzy, i.e. they have no sharp limits. This is expected because of the continuous nature of the parameters, and because of both uncertainties and measurement errors. This is an important point that is essentially overlooked by the other methods, and that should be kept in mind. 

As we see below, these four groups are easily identified with the groups obtained by cladistics, and for this reason they are not given labels in this paper.

\subsection{Cladistic analysis}
\label{classClad}

%%%%%%%%%%%
   \begin{figure}
   \centering
 \includegraphics[width=\columnwidth,angle=0]{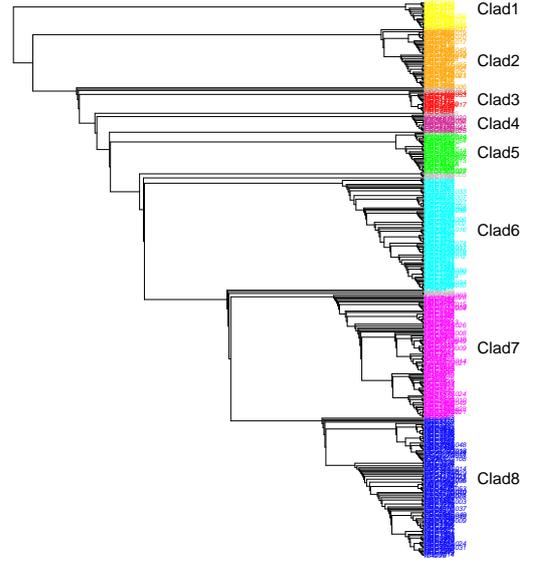}
   \caption{Most parsimonious tree found with cladistics with the identification of the eight groups and their corresponding colours.} 
    \label{figtreeclad}%
    \end{figure}
%%%%%%%%%%%%%

The cladistic analysis, performed with the six parameters selected in Sect.~\ref{parconcl}, produces a most parsimonious tree (shown in Fig.~\ref{figtreeclad}), on which we can identify groups. There is no absolute rule for defining groups on a cladogram. However, substructures in the tree are a good guide.

We identified eight groups in this tree, three large ones with more than 80 objects, an intermediate group with about 50 objects, and four smaller ones with fewer than 30 members (Fig.~\ref{figtreeclad}). These groups are named ``Clad1'' (the most ancestral one, at the top) to ``Clad8'' (at the bottom) 
(Fig.~\ref{figtreeclad}). This numbering and presentation of the tree should not a priori be seen as a diversification arrow since branches can be switched graphically. 
It is the physical interpretation that both confirms the possible ancestrality of group Clad1 and gives the right order of diversification. 

The rooting of the tree (i.e. the choice of objects that appear graphically at the top of the tree and are supposed to be the closest to the common ``ancestor species'' of all the objects of the sample) is necessary to define the direction of diversification, and in general affects the contours of the groups. The tree is here rooted with the group of the lowest average metallicity as measured by \MgbFe\ according to our assumption that the lowest metallicity corresponds to the most  ancestral objects. This guess is monovariate and might not represent the best choice in a multivariate study like the present one. However, we do not yet have a better multivariate criterion for primitiveness. The rooting of the tree can easily be changed.

In contrast to the other partitioning methods, some objects appear to be isolated on the tree and consequently cannot be easily grouped with others. Each of them could indeed represent a class, but, for the sake of simplicity, we decided not to identify them with specific colours. We simply gather all such objects as Clad0, give them a grey colour in plots or simply disregard them in the discussions that deal with the statistical properties of groups.

\subsection{Comparison of the four partitionings}
\label{classcomp}

The four methods produce three (PCA+CA), four (MCA), seven (cluster analysis), and eight (cladistics) groups. They thus all agree for a relatively small number of groups.

The agreement between cladistics and PCA+CA is quite good (see Fig.~\ref{figcomppart}), if we identify the three following groups: (Clad1,Clad2,Clad3,Clad4,Clad5, and a part of Clad6), (Clad7), and (Clad6,Clad8) with PCACA1, PCACA2, and PCACA3 respectively.

The agreement between the PCA+CA and cluster analyses is also quite good with PCACA3 being composed of Clus4, Clus7, and a part of Clus6, PCACA2 being composed essentially of Clus5 and partly Clus6, and PCACA1 being mainly composed of Clus1, Clus3, and part of Clus4.

Cluster and cladistic partitionings agree very well for the three large groups (Clus4$\simeq$Clad6, Clus5+Clus6$\simeq$Clad7, and Clus7$\simeq$Clad8). The situation is slightly more complicated for the other groups, but still convergent with Clus1$\simeq$Clad1+Clad3, and Clad2 being split mainly into Clus2, Clus3, and also Clus4. Conversely, Clus2 is contained mainly in Clad2 and also in Clad7.

The four groups from MCA are in very good correspondence with groups Clad6, Clad7, Clad8, and Clad1+Clad3. 
On the other hand, Clad2+Clad5 does not seem well-justified based on the MCA result. Interestingly, Clad6 and Clad8 are not quite independent, in agreement with PCACA3 being mainly composed of Clad6 and Clad8 as seen above.

The PCA+CA identifies a smaller number of groups than the other partitionings do. 
This was expected, because of the effect of the PCA analysis, which eliminates too many of the correlations (Sect.~\ref{metPCA}). The MCA result reinforces the significance of groups Clad6, Clad7, Clad8, and Clad1+Clad3, which are all also identified in the cluster analysis. The other groups from cladistics and cluster analyses are either less robust or more fuzzy.

We conclude that the number of groups is at least four, and probably seven or eight. In the following, we consider the cladistic result with eight groups because it provides the very important evolutionary relationships between them.

Supplementary figures are given in Appendix~\ref{appendFigures} for the cluster partitioning and can be used to check that our interpretation does not depend on the detailed boundaries of the groups. In addition, two complementary cladistic analyses were performed in order to check the influences of two different determinations of the distance of galaxies (needed to determine $r_e$), and of measurement errors are presented in Appendix~\ref{fourparam}.

\section{Discriminant descriptors of galaxies}
\label{descriptors}

Among the initial 23 quantitative parameters (Sect.~\ref{data}), only 6 are discriminant and actually yield a relatively robust partitioning (Sect.~\ref{parconcl}).  The 17 remaining parameters do not yield enough information to distinguish different classes of objects, because either intrinsically they are not informative, they bear the same redundant information as the discriminant ones, or they are not discriminant for the sample under study.

It is remarkable that the global luminosity of the galaxies (\Mabs\ or \Kabs) is not discriminant. It is usually used as an indicator of mass and chosen as a main criterion of a priori classification. Luminosity is also often assumed to characterize the level of evolution (for instance in the so-called ``downsizing effect''). However, from a diversification point of view, the absence of the global luminosity is expected, since mass is a global property that can be acquired by different processes, i.e. accretion or merging, which have different timescales and perturbing powers. Such parameters, which show too much convergence, are not well-suited to establish phylogenies, that is, they are not good tracers of the assembly history of galaxies. Mass is bound to increase, it is thus not specific to any particular assembly history, which could distinguish different kinds of galaxies. Nevertheless, mass is not entirely absent and is represented somehow in \logs\ and \Brie, which are certainly better tracers  than 
mass itself of the way in which mass has been assembled.

The index \OIII, which tends to decrease in more metallic galaxies, is a discriminant parameter, but \Hbeta, which is often used as an age indicator, is not. This is not so surprising since age is not an indicator of diversity, as it is shared by all objects \citep[see a discussion in][]{FDC09}. Age, even more so than either mass or size, is bound to increase independently of the assembly history. Anyhow, defining an age for a galaxy is tricky and is often taken as the average age of the stellar populations, which is a poor tracer of the assembly history. 

The size parameters \logre\ and \ldiam\ are not discriminant. They are probably merely scaling factors that are somewhat similar to mass, and bound to increase regardless of the sequence of transforming events that occur during the assembly history of galaxies. However size does not seem to be represented at all, and if so probably weakly in \logs\ and \Brie, or even in some hidden correlation, which we study later on in this paper. 

However, one may wonder why \citet{Fraix2010} found a robust partitioning using only four parameters. Two of them are in our list of six (\logs\ and \Brie), one of them ($Mg_2$) is very similar to \MgbFe, but the fourth one is \logre, which is not a discriminant parameter in the present analysis. There are several reasons for this.

First, in \citet{Fraix2010}, the four parameters were not the result of a multivariate and objective selection, but were chosen because common wisdom suggests that they may be important for characterizing the physics of galaxies. The very positive result obtained with these four parameters strongly supports this a priori, but the present paper demonstrates that only three of them are really discriminant parameters.

Second, three parameters out of four are discriminant, so that the partitioning signal is borne by these three. Unless the fourth parameter (\logre) is strongly erratic or contradictory, this signal is not expected to be entirely destroyed (see Sect.~\ref{fundplane}). 

Third, the discriminant parameters may in principle be different from one sample to another, if the diversity of objects is not equally covered. They may also depend on the initial set of parameters, if more discriminant ones are present in a larger list. This is probably the case for \logre, which has been replaced by better observables.

It is thus unsurprising that the four parameters used in \citet{Fraix2010} yielded a robust partitioning and that we find more discriminant parameters in the present study. The six parameters selected in the present analyses are not necessarily the most suitable ones for other samples, for which new partitioning analyses should ideally be conducted.

\section{Group properties}
\label{discussion}

%%%%%%%%%%%
   \begin{figure}
   \centering
 \includegraphics[width=\columnwidth]{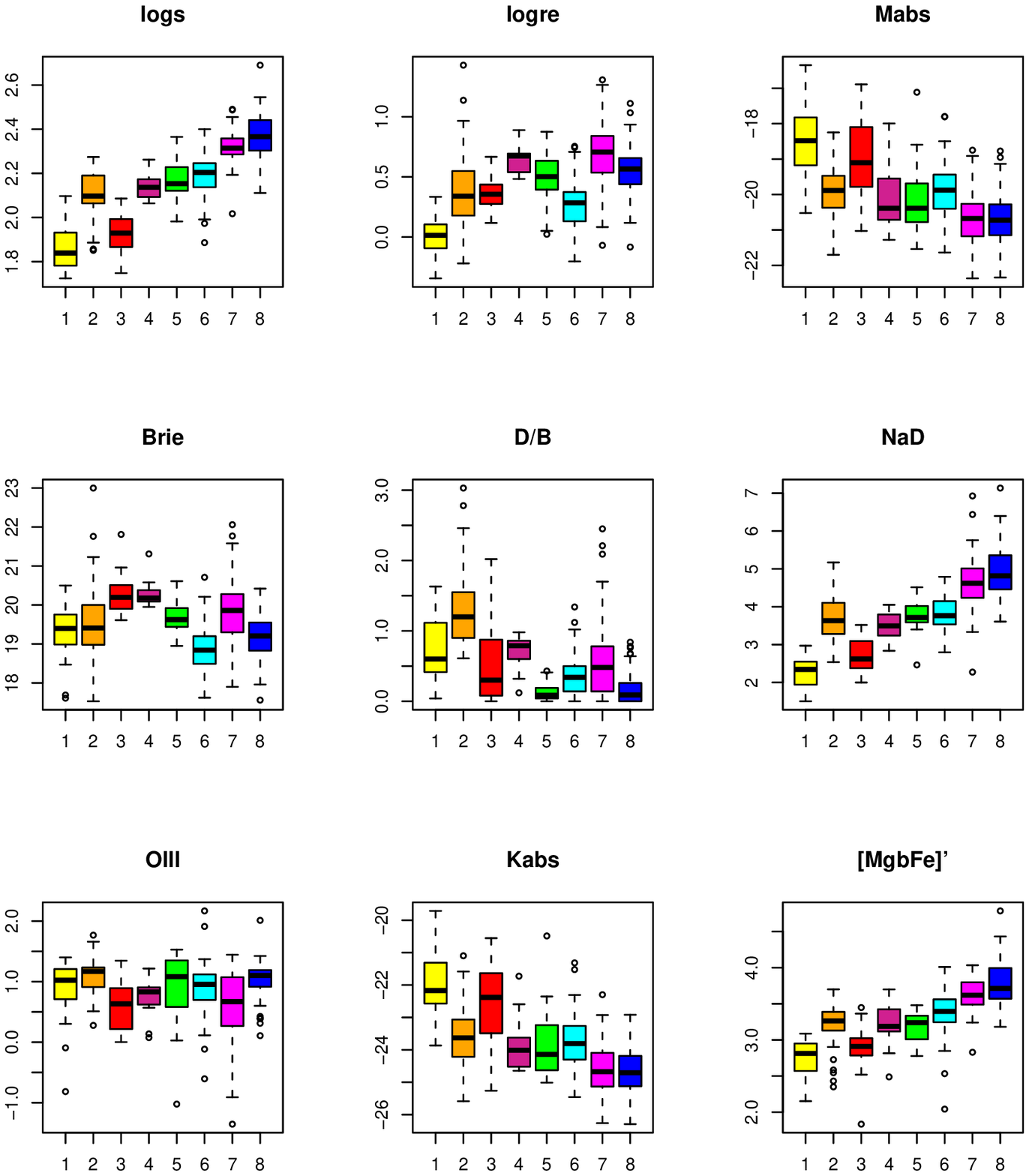}
 \includegraphics[width=\columnwidth]{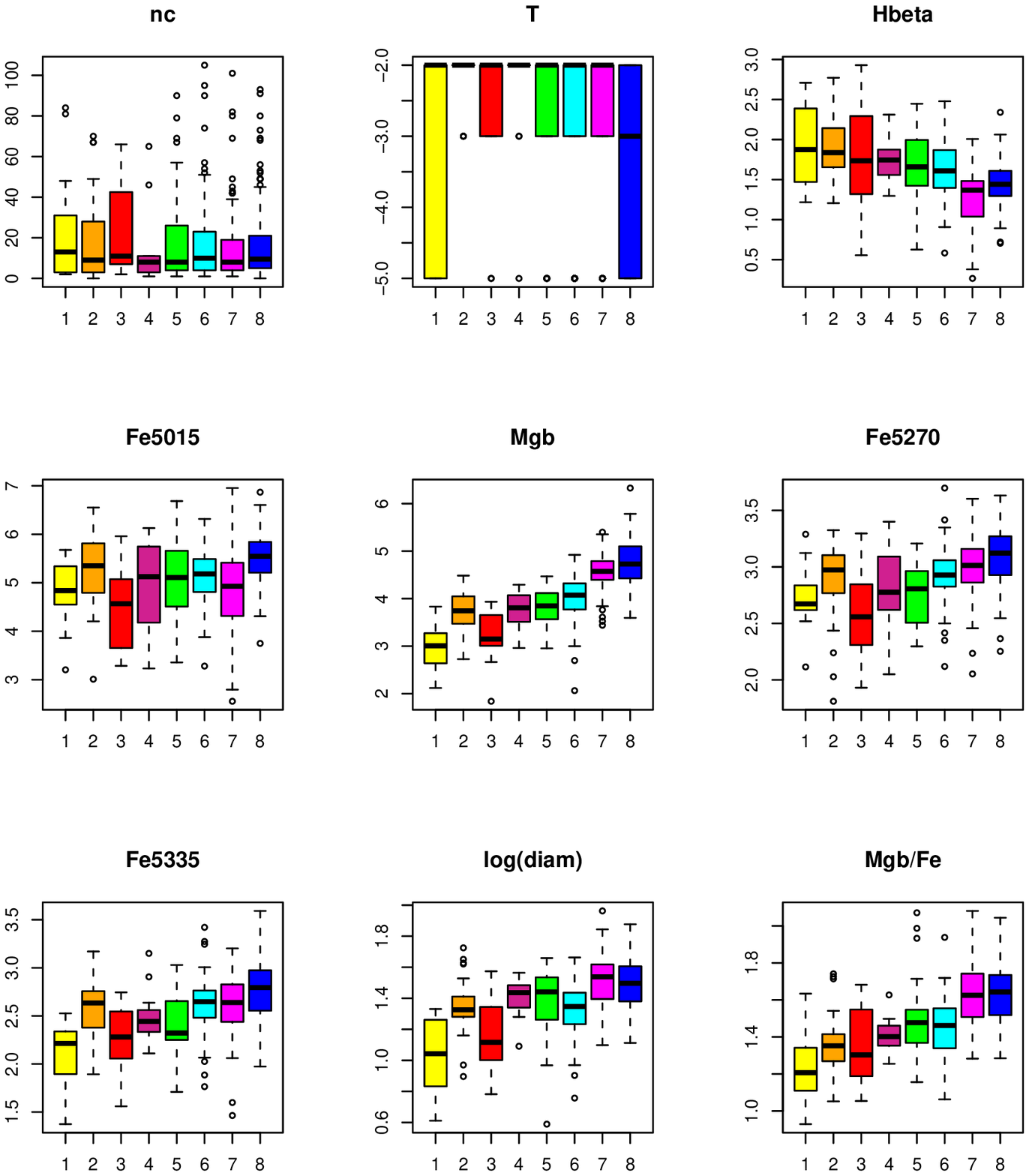}
 \includegraphics[width=4 true cm]{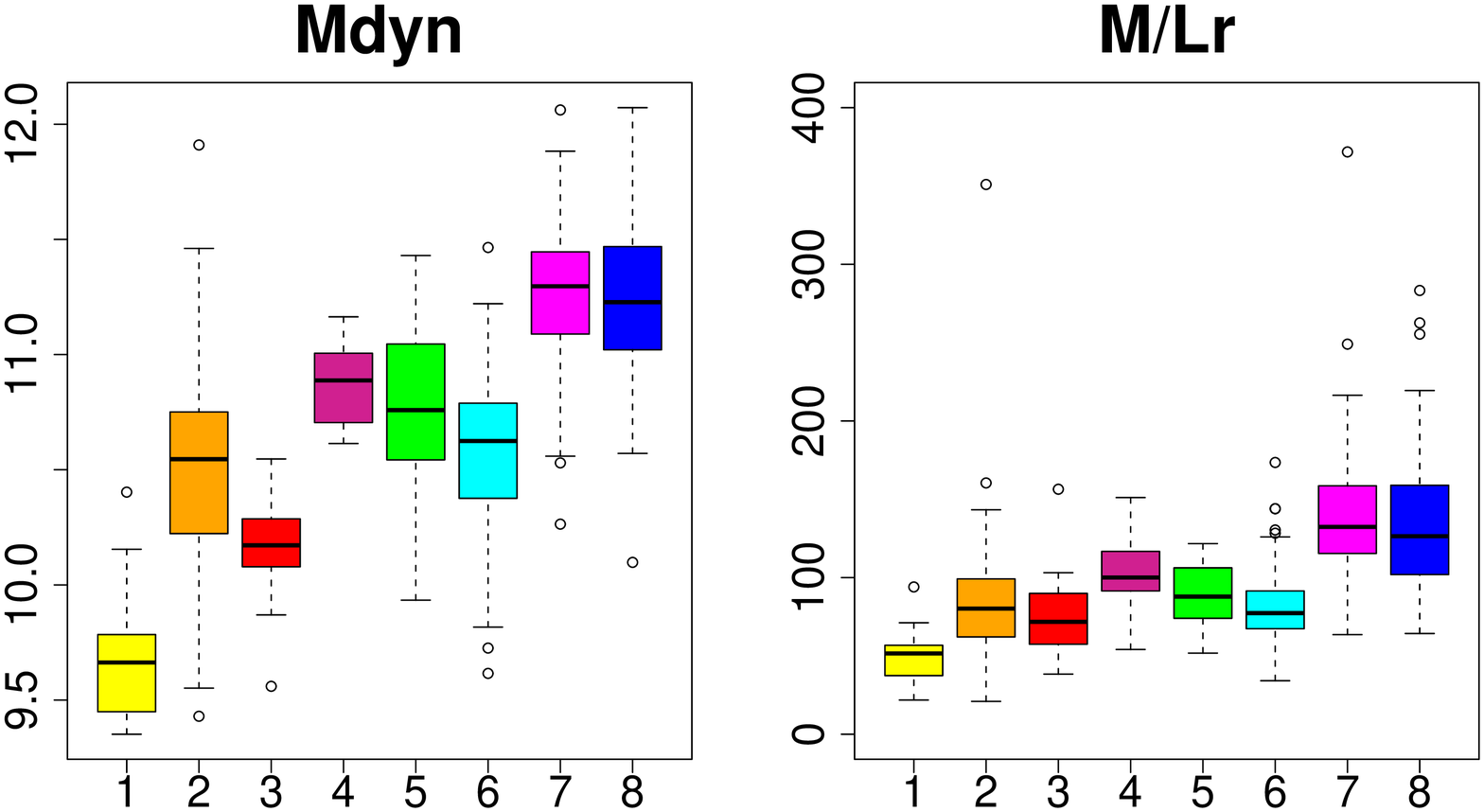}
   \caption{Selection of the most interesting boxplots.} 
    \label{figboxplotClad}%
    \end{figure}
%%%%%%%%%%%%%

%%%%%%%%%%%
   \begin{figure}
   \centering
 \includegraphics[width=\columnwidth]{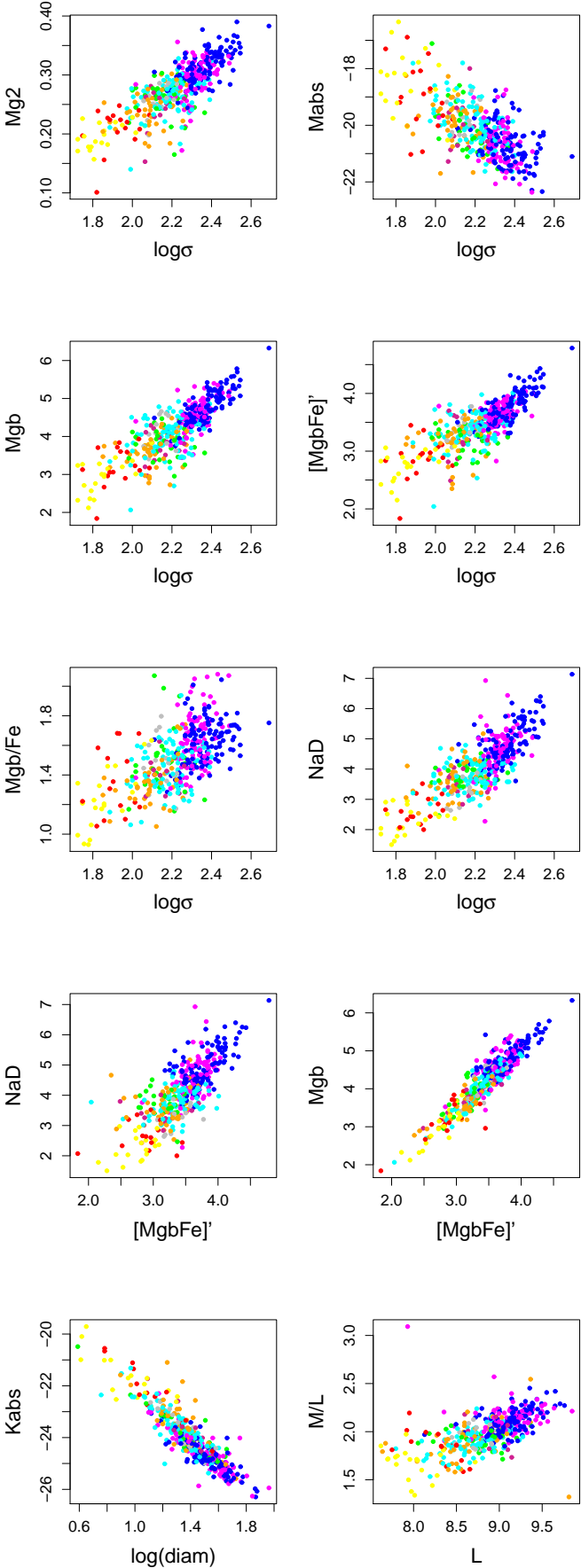}
   \caption{Scatter plots showing evolutionary correlations. Colours are the same as in Fig.~\ref{figboxplotClad} and Fig.~\ref{figtreeclad}.} 
    \label{figspurious}%
    \end{figure}
%%%%%%%%%%%%%

The groups identified by the partitioning methods must be understood in the light of their statistical and comparative properties. In this section, we first identify the main trends along diversification and then describe the distinctive properties of the groups.

For this purpose, we use boxplots, which give the four quantiles of each parameter for the eight groups. We consider two additional parameters. The dynamical mass  is defined as $M_{dyn}\simeq A \sigma^{2}R_{e}/G$ with $A=3.8$ (with $M_{dyn}$ in solar mass, $\sigma$ in km s$^{-1}$ and $R_e$ in kpc) according to \citet{Hopkins2008}, as done in \citet{Fraix2010}. This makes $M_{dyn}\simeq 5.95*\sigma^{2}R_{e}$. Using this mass, we compute the mass-to-light ratio $M/L_r$ using \Brie$ = -2.5\log(L_r /\pi r_e^2)+4.29$. 

We show the most informative boxplots in Fig.~\ref{figboxplotClad}; the others do not indicate significant differences between groups. We show the boxplots for the cluster partitioning in Fig.~\ref{figboxplotClus} of Appendix~\ref{appendFigures}.

Figure~\ref{figboxplotClad} shows that \logs, \logre, \NaD, \MgbFe, \Mgb, \ldiam, $Fe5270$, $Fe5335$, \MgbsFe, $M_{dyn}$, and $M/L_r$ essentially increase along the diversification rank defined on the tree of Fig.~\ref{figtreeclad}, while  \Hbeta, \Mabs, \Kabs, and possibly \DsB\ decrease. As already mentioned (Sect.~\ref{classClad}), this
rank is not necessarily as linear as it seems. Anyhow, the adopted rooting of the tree gives a very sensible result: globally, galaxies tend to become more metallic, more luminous, more massive, and larger with increasing diversification. At the same time, they acquire a larger central velocity dispersion, which is often related to the higher mass, and \NaD\ is also known to increase with both mass and velocity dispersion. In addition, the decrease in \Hbeta\ indicates that the average age of the stellar populations increases with diversification.

\MgbsFe\ increases with diversification. The index [$\alpha$/Fe] is known to increase with galaxy mass and age, because successive mergers and accretions trigger more intense star formation on shorter timescales. These events clearly participate in the diversification of galaxies, confirming our observed increase in \MgbsFe.

The \OIII\ index does not show any trend with diversification, but has a lower median value for the three groups Clad3, Clad4, and Clad7. 

There is no systematic trend in environmental properties with diversification, $n_c$ having a wide range in all groups, except in Clad4 where it is small. Since an observed galaxy is the result of a long and multiple sequence of transforming events, 
it is probably the past environment, rather than the observed one, that plays a role in the diversification process.

The most diversified groups (Clad5 to Clad8) have on average a lower \DsB\ ratio, suggesting that transforming events, such as accretion, interaction, and mergers, tend to destroy discs and build larger bulges, presumably by randomizing stellar orbits. 

The morphological type is unevenly distributed among groups, Clad2 and Clad 4 having nearly only galaxies with $T=-2$, whereas $T=-5$ galaxies are found mainly in Clad1 and Clad8 (respectively, the most ancestral and one of the most diversified groups).

Apart from the general trends with diversification, the groups have distinctive properties, otherwise there would be no reason for finding separate groups. Their distinctive properties are the following, where the number of members is given in parentheses:

\begin{itemize}
 \item Clad 1 (20 objects): these galaxies have the properties expected from an ancestral group in being small, faint, discy, and of low metallicity. They are young (although have a large spread in \Hbeta) and have a large spread in morphological type.  They have very low $M_{dyn}$\ and \logs, and a low $M/L_r$. 
 \item Clad2 (45): these galaxies have the same average \Hbeta, \Brie, and \OIII\ as Clad1. They are larger, brighter, more massive, more metallic, and have a much higher \logs\ than both Clad1 and Clad3. They are all of morphological type $T=-2$, and have the highest \DsB\ of all groups by far.  
 \item Clad3 (16): these galaxies have a large range in many parameters (\Hbeta, \Mabs\ and \Kabs, \ldiam, $n_c$, \OIII, \MgbsFe), but not in \MgbFe, \logre, \logs, \Brie, \NaD, or $M_{dyn}$. They have a small \ldiam\ similar to Clad1, but a much higher \logre. They are relatively faint with a relatively low \logs\ and \OIII, and a high \Brie.
 \item Clad4 (13): these galaxies resemble those of Clad2 in most respects (see discussion below on the respective placements of Clad2 and Clad3). In particular, they are all of morphological type $T=-2$. The main differences are that Clad4 objects are large (the highest \logre\ after Clad7), of low surface brightness (high \Brie) , have higher $M_{dyn}$ and $M/L_r$, and are slightly less discy.
 \item Clad5 (30): these galaxies are very similar to the Clad4 ones, except that they have a low \Brie\ and a very low \DsB, the lowest of all groups with Clad8. 
 \item Clad6 (85): this is one of the three largest groups, which are also the most diversified. Its galaxies have unexpectedly low values of $M_{dyn}$, \logre, and \Brie. Interestingly, they have very similar properties to those of Clad2 galaxies, except for a much lower \DsB, slightly lower \Brie, and \Hbeta, and a slighty higher \MgbsFe.
 \item Clad7 (94): these galaxies are the largest in this sample. They are the most luminous and have the highest \logs, $M_{dyn}$, and $M/L_r$\  together with Clad8. Clad7 galaxies have a higher \Brie, slightly lower \Hbeta\ and \OIII, and a slightly higher \DsB\ than Clad6 and Clad8.
 \item Clad8 (106): their distinctive properties are a very low \DsB\ and a very large spread in morphological type, in a similar way to Clad1. They have a higher \logs\ and a lower \Brie\ than galaxies of Clad7. They have values of $M_{dyn}$\ and $M/L_r$ that are as high as Clad7.
\end{itemize}

%%%%%%%%%%%
   \begin{figure*}
   \centering
 \includegraphics[width=\columnwidth]{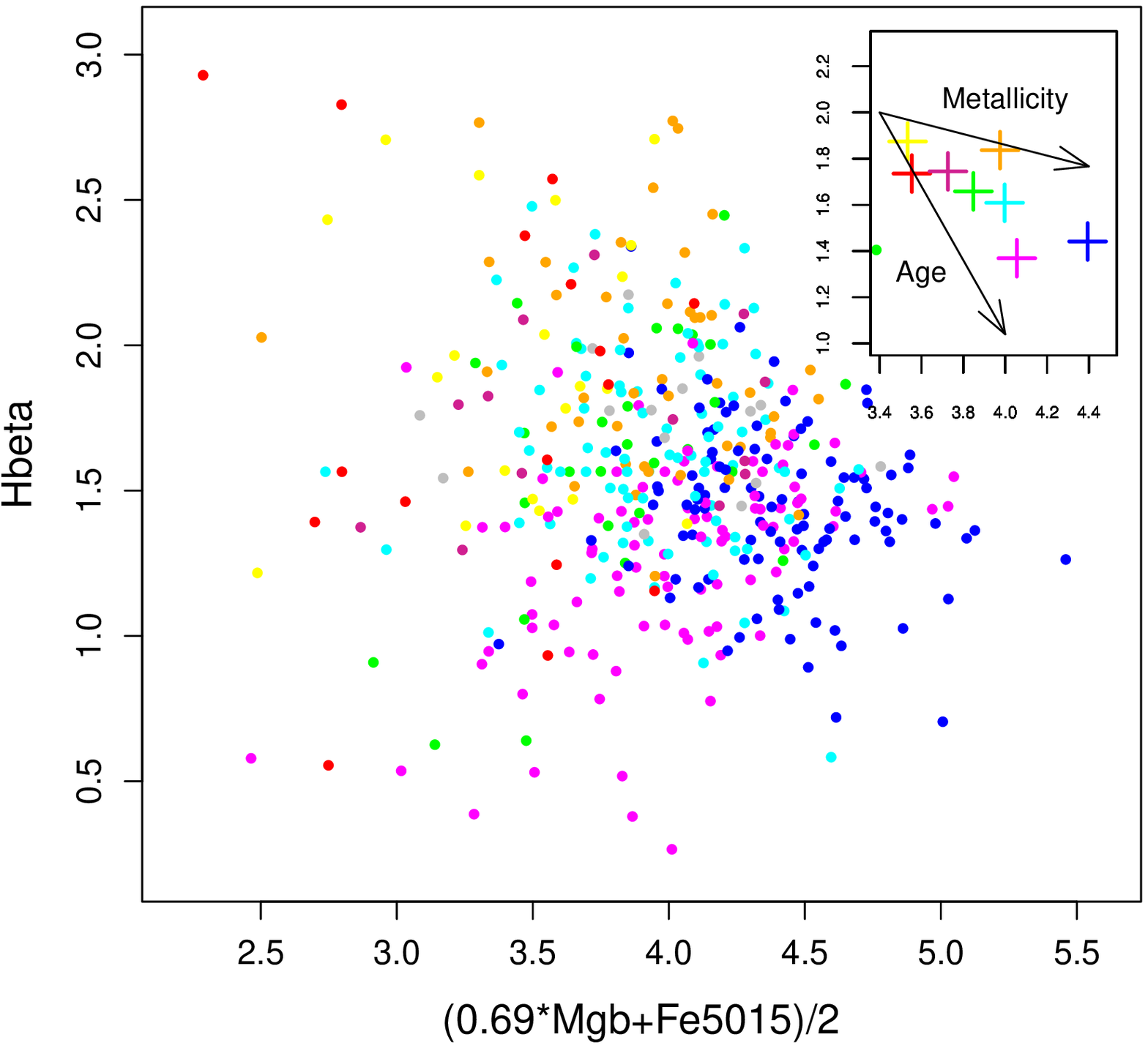}
 \includegraphics[width=\columnwidth]{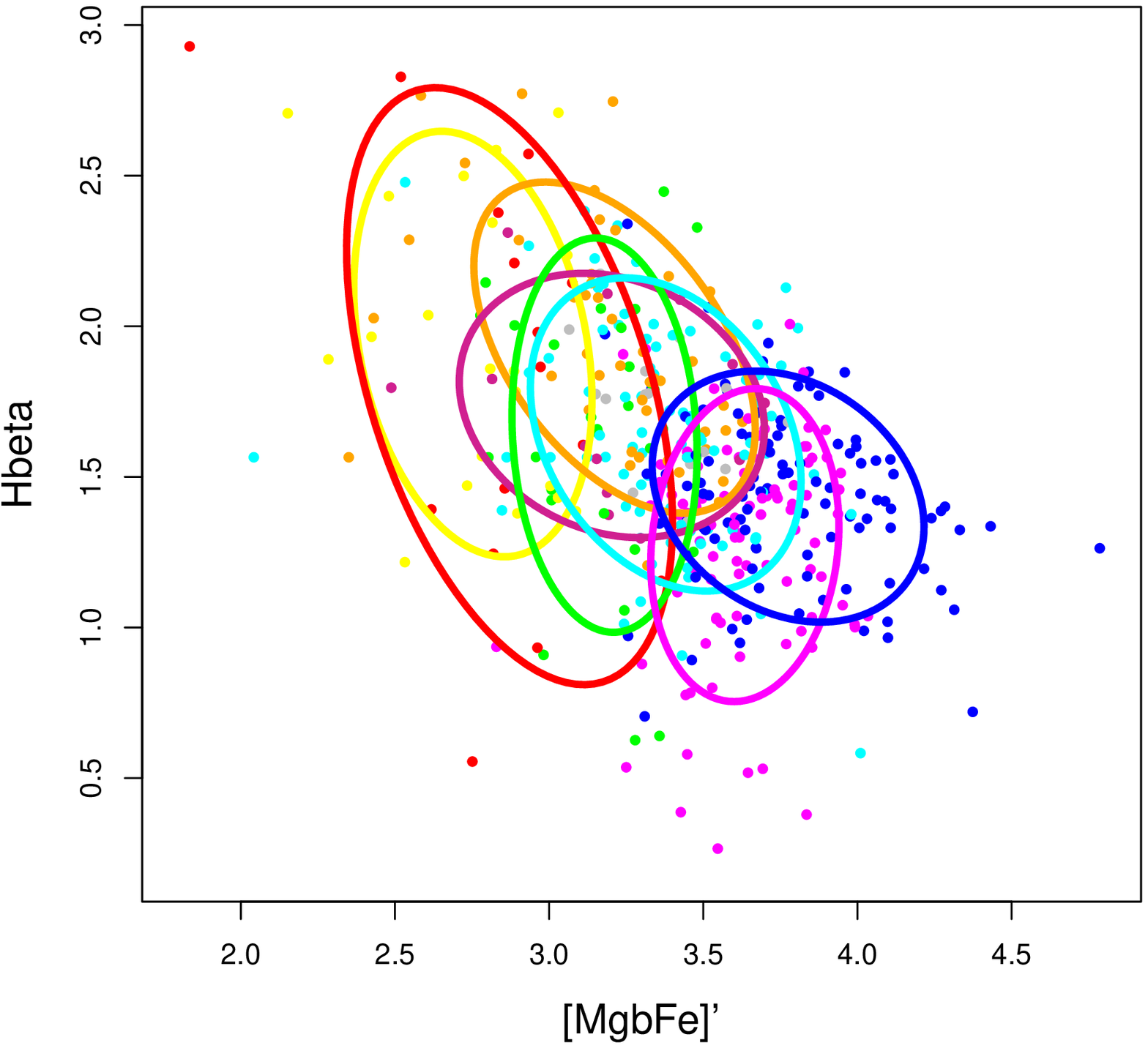}
   \caption{Left: Equivalent of Figure 9 by \citet{Thomas2010} or Figures 3 and 8 by \citet{Kuntschner2010}. The inset shows the median for each group and the arrows indicating the direction of increase of age and metallicity from single stellar population evolutionary models as shown by these authors. Right: Group inertia ellipses for the \Hbeta\ vs \MgbFe\ scatter plot.} 
    \label{figellipses}%
    \end{figure*}
%%%%%%%%%%%%%

The Clad2 group often departs from the general trend along diversification (Fig.~\ref{figboxplotClad}), which would seem smoother if Clad2 and Clad3 were inverted. We have noted (Sect.~\ref{classcomp}) that Clad2 is split between Clus2, Clus3, and also Clus4. It is significantly higher than expected in \logs, \DsB, \NaD, \MgbFe, \Mgb, \ldiam,
$Fe5270$, $Fe5335$, and lower in \Mabs\ and \Kabs. This means that, because of some parameters, it seems misplaced in the diversification scenario for other parameters. This behaviour is visible in the partitioning from the cluster analysis (Fig.~\ref{figboxplotClus}) since Clus2 and Clad2 are partially similar. Hence, why was Clad2 placed so early by the cladistic analysis, while it is more diversified in four of the six parameters used for the analysis? 

The diversification scenario given in the tree of Fig.~\ref{figtreeclad} is obtained from the parsimony criterion, which chooses the simplest combined evolution of all parameters. Taken individually, the simplest evolutionary curve of each variable is monotonic with as few reversals as possible. For instance, in the \logs\ boxplot of Fig.~\ref{figboxplotClad}, one would expect Clad2 and Clad3 to be inverted to avoid the Clad2 box to ``peak''. However, this is a multivariate compromise, and since Clad2 would be better placed in the very first position on the \DsB\ plot, to ``smooth'' the evolutionary curve, it is understandable that this is the most parsimonious placement on the tree. 
In addition, while the two discriminant parameters \Brie\ and \OIII\ show a variable behaviour, they would nevertheless induce us to place Clad2 before Clad3. 

The conclusion is that Clad2 is correctly placed in second position, because this is a multivariate analysis, which seeks a compromise among several parameters. This shows the importance of selecting the parameters objectively, with multivariate tools. Otherwise, with too many redundant parameters, the peculiar properties of Clad2 could have easily been lost.

The relative and distinctive properties of the galaxies from the different groups obviously cannot be summarized with only one or two physical parameters. The relative properties of the groups show that the evolution of galaxies is not linear. The global trend in some properties (such as mass, metallicity, or \Hbeta) may appear to be roughly linear globally, but a detailed analysis, and especially the distinctive properties within each group, give many clues to understand the assembly history of the corresponding galaxies.

Having highlighted the group properties, we examine the possible correlations between them in the next two sections.

\section{Scatter plots and correlations}
\label{scattercorrel}

%%%%%%%%%%%
   \begin{figure}
   \centering
 \includegraphics[width=\columnwidth]{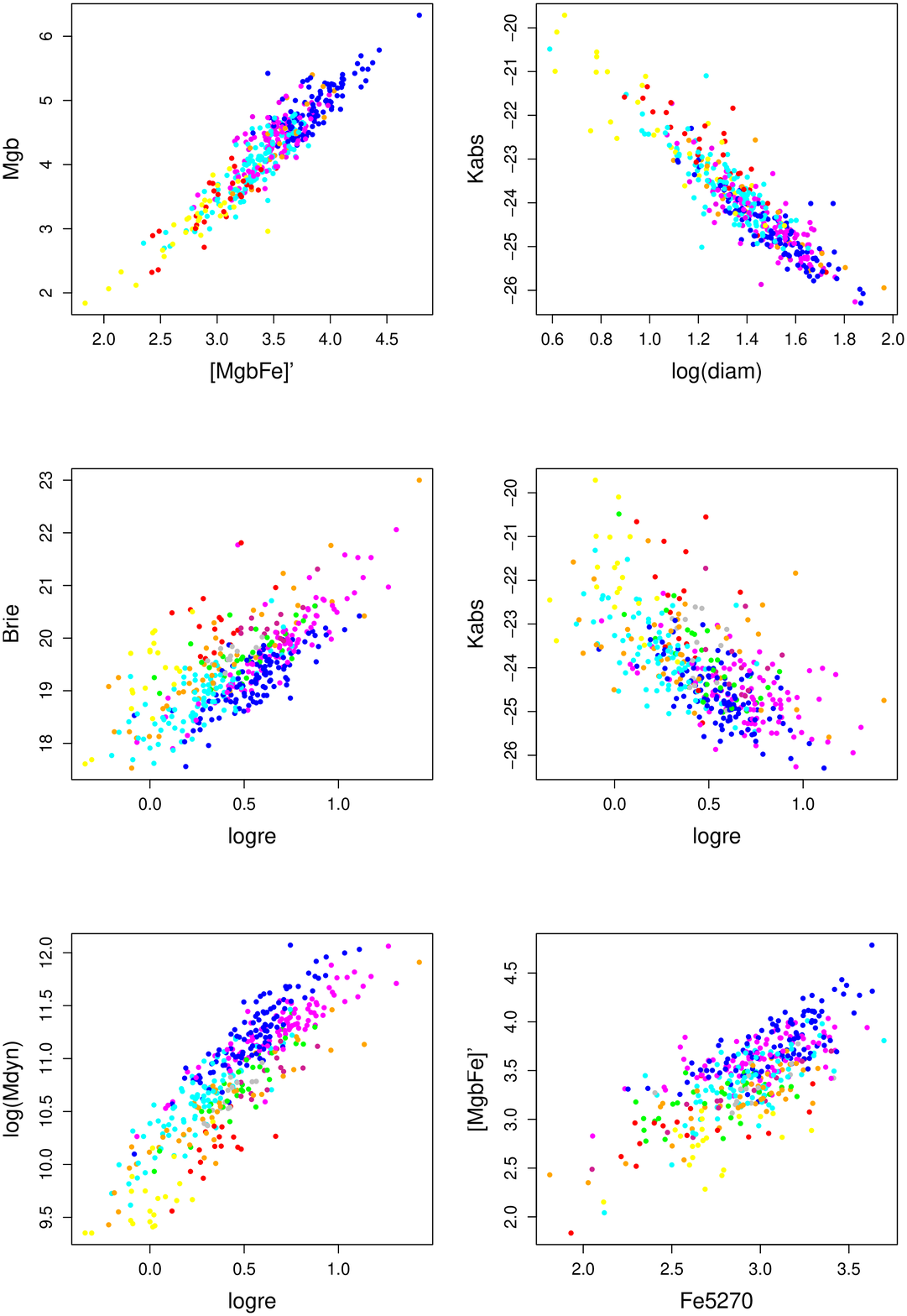}
   \caption{Scatter plots showing correlations within groups.} 
    \label{figsgroupdep}%
    \end{figure}
%%%%%%%%%%%%%

Scatter plots must be examined using the partitioning to look for different behaviours between groups or within groups. 

In the first case, the distribution of the groups traces the projected evolutionary track given by the tree. The fundamental plane is one example \citep[Sect.~\ref{fundplane} and][]{Fraix2010}. We however focus on cases showing a roughly linear track, where the groups are approximatively ordered along a linear correlation. We refer to these relations as evolutionary correlations (Sect.~\ref{corrspur}) since these groups are related in cladistics by evolutionary relationships. They are important since they imply that the observed relation can be generated mainly by evolution, as found in \citet[][]{Fraix2010} and formalised in \citet{DFB2011}.

In the second case (Sect.~\ref{corrcausal}), some correlation may or may not be present within a given group, independently of the global behaviour between groups.

\subsection{Evolutionary correlations}
\label{corrspur}

We confirm the evolutionary nature of the $Mg_2$ -- \logs\  correlation found by \citet{Fraix2010} and identify several other cases, the clearest ones being shown in Fig.~\ref{figspurious}. Such evolutionary correlations are revealed by the succession of groups ordered along the correlation with the most ancestral group (Clad1) at one end and the most diversified ones (Clad7 and Clad8 here) at the other end. 

Several of these evolutionary correlations involve the following set of parameters: \logs, \logre, \Mabs\ (and \Kabs), \Hbeta, \Mgb\ (and $Mg1$, $Mg_2$, \MgbFe, and \MgbsFe), \NaD, and \ldiam. Some relations are particularly tight (such as \ldiam\ vs \Kabs or \Mgb\ vs \MgbFe). The iron Lick indices $Fe5270$, $Fe5335$, and $Fe5406$, as well as \HmK, also follow an evolutionary correlation with \Mabs, although it is quite loose.

In all cases, except for \ldiam\ vs \Kabs\ and \Mgb\ vs \MgbFe\  (which are discussed in Sect.~\ref{corrcausal}), 
the correlation is not present within each group. This is a clear sign that there is no direct causal physical link between the two variables, but simply a change on average with galaxy diversification.

\citet{Thomas2005} discuss the origin of the \Mgb\ vs \logs\ correlation. They find that metallicity, not age, is the main driving factor. This would again justify the use of metallicity as a reasonable tracer of diversification. However, we find instead that the \Mgb\ vs \logs\ correlation is an evolutionary correlation, implying that diversification is indeed the real driver: metallicity, like central velocity dispersion, is bound to change on average as the galaxies evolve. This could explain why investigations find that this correlation appears so sensitive to several parameters: it has been proposed to be driven by metallicity, age, and relative abundance of different heavy elements \citep[see][for references]{Matkovic2009}. This sensitivity more probably points to an underlying, hidden, and confounding factor, which creates the apparent correlation \citep{DFB2011}.

The correlations between \Mgb\ and \MgbFe\ and between \NaD\ and \MgbFe\ are clearly evolutionary, with diversification increasing from left to right in Fig.~\ref{figspurious}, while the correlations found by \citet{Thomas2010} are driven by total metallicity, which, for a given age and light-element ratio, increases from left to right in their Figs~6 and 8. The dispersion in the \NaD\ vs \MgbFe\ relation is larger than
that in the \Mgb\ vs \MgbFe\ relation for our data and theirs, and not well-accounted for by
their model, presumably because of the fixed age and light-element ratio assumptions. Nevertheless, there is agreement between our result and their model since the average metallicity of galaxies obviously increases with diversification.

The \MgbFe\ vs \logs, \MgbsFe\ vs \logs, and \Mgb\ vs \logs\ correlations (Fig.~\ref{figspurious}) can be compared with the $Z/H$ vs \logs\ and $\alpha/H$ vs \logs\ in Figure 16 in \citet{Kuntschner2010}. The same correlation is present, but we clearly show its evolutionary nature. \citet{Kuntschner2010} ask the question: ``\textit{What drives the [$\alpha$/Fe] -- $\log\sigma_e$ (or mass) relation?}''. Our answer is simply: diversification. They indeed arrive at the same conclusion because they find that, in their sample, ``\textit{there is evidence that the young stars with more solar-like [$\alpha$/Fe] ratios, created in fast-rotating disc-like components in low- and intermediate-mass galaxies, reduce the global [$\alpha$/Fe] and thus significantly contribute to the apparent [$\alpha$/Fe] -- $\log\sigma_e$ relation}''. These galaxies belong to our Clad1 group as stated above, but they are not the sole responsible cause of this ``apparent'' relation, since all our groups are aligned along the same trend.

The Faber-Jackson relation, \Mabs\ vs \logs\ (Fig.~\ref{figspurious}), also appears to be a purely evolutionary correlation: the sequence of evolutionary groups are aligned along this correlation. If mass had been the hidden parameter, then a similar correlation should exist within each group. This is not the case. This result was corroborated in an independent way by \citet{NigocheNetro2011}.

%%%%%%%%%%%
   \begin{figure}
   \centering
 \includegraphics[width=\columnwidth]{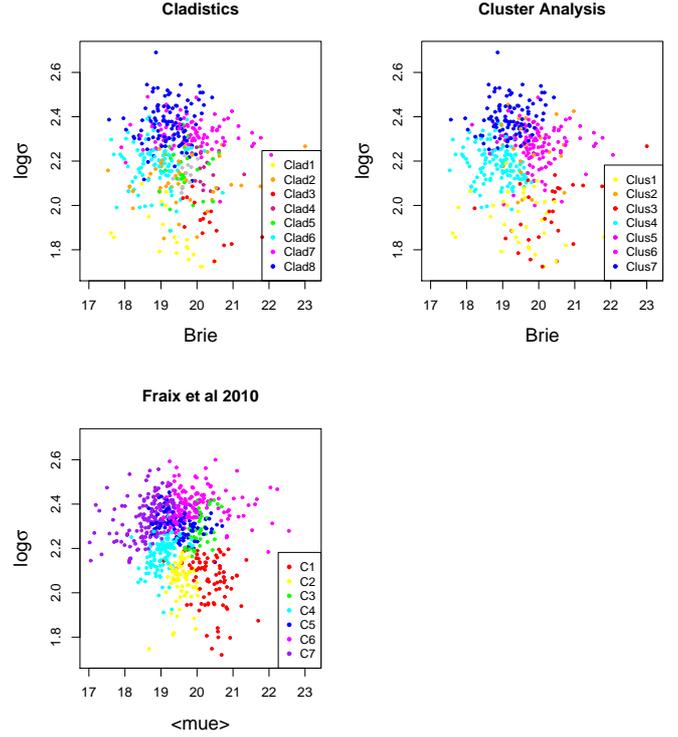}
   \caption{Projection on the fundamental plane. Comparison between the partitionings obtained by cluster and cladistic analyses and that of \citet{Fraix2010}. See also FIg.\ref{figcompgroupsFP}}.
    \label{figfundplane}%
    \end{figure}
%%%%%%%%%%%%%

%%%%%%%%%%%
   \begin{figure}
   \centering
 \includegraphics[width=\columnwidth]{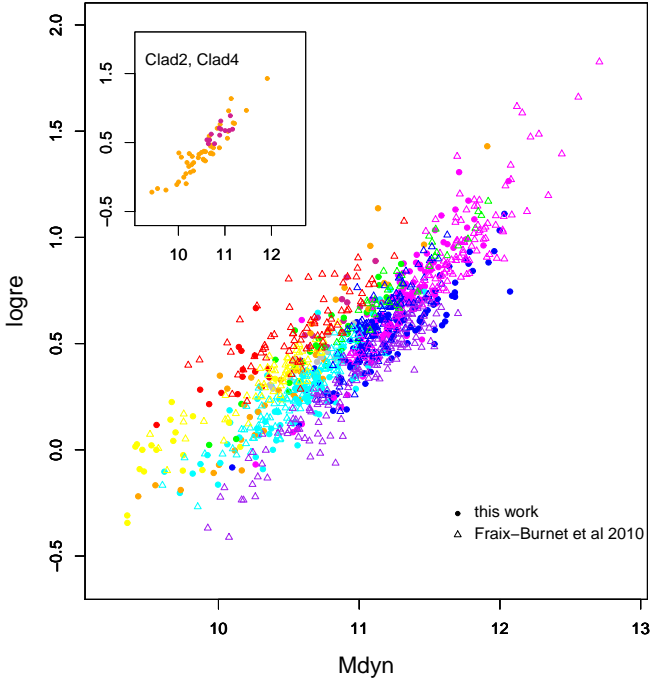}
   \caption{Comparison between the partitioning obtained by cladistics and that of \citet{Fraix2010}. In the inset, we plot the groups Clad2 and Clad4, which are difficult to see in the main graph. See also Fig.\ref{figcompgroupslogreMdyn}.} 
    \label{figcompMdynRe}%
    \end{figure}
%%%%%%%%%%%%%

\subsection{Diversification or ageing?}
\label{metagedegen}

There is a well-known degeneracy between the age (measured by \Hbeta) and the metallicity (measured by \MgbFe\ or $(0.69*Mgb + Fe5015)/2$) of stellar populations, which models of stellar evolution have tried to circumvent \citep[e.g.][]{TripiccoBell1995}. In particular, \citet{Thomas2010} and \citet{Kuntschner2010} superimposed evolutionary tracks from different models on galaxy observations of \Hbeta\ as a function of a metallicity indicator defined by $(0.69*Mgb + Fe5015)/2$. We reproduce the same figures in Fig.~\ref{figellipses} (left),
showing with crosses the median for each group and with arrows the principal direction of increases in age and metallicity. We emphasize that the stellar evolution models used by \citet{Thomas2010} and \citet{Kuntschner2010} to produce their figures 
are single population models with fixed solar value of [$\alpha$/Fe], while the data used in this paper (Sect.~\ref{data}) are integrated over the whole galaxy, mixing together the contributions of possibly several different stellar populations. 

Our groups are clearly arranged according to diversification following an increase in both age and metallicity. The spread of the correlation is large, and, within each individual group, the range in age and metallicity is also large. This dispersion could certainly be explained by many factors, such as an extended horizontal branch, which can increase \Hbeta\ \citep[e.g.][]{Greggio1997,Matkovic2009}. In all cases, the median for each group is nearly perfectly aligned between the two axes for age and metallicity, ordered as in Fig.~\ref{figtreeclad}, except for Clad2 and Clad7, which depart from the main alignment. This kind of plot indeed merely tells us that the age, metallicity, and \Hbeta\ of galaxies evolve on average with time.  

However, we find no correlation between age and metallicity {\it within} individual groups. This is clearly shown in Fig.~\ref{figellipses} (right), which plots \Hbeta\ vs 
\MgbFe, which is a better indicator of metallicity than \Mgb. It is striking that the elongated inertial ellipses for Clad3, Clad5, and Clad7 are well-aligned with the \Hbeta\ axis, with relatively little spread in metallicity, the one for Clad1 is slightly inclined, and the one for Clad2 is aligned along the global trend. Since the other ellipses are quite round, Clad1 and Clad2 are the sole groups that might show a barely significant correlation between \Hbeta\ and a metallicity indicator.

The wider range in \Hbeta\ for the less diversified groups, especially Clad1 and Clad3, clearly appears in Fig.~\ref{figellipses} and can also be seen in Fig.~\ref{figboxplotClad}. It probably corresponds to the well-known wider range in age of the low-mass objects \citep[e.g.][]{Matkovic2009}. Our interpretation is not that the low-mass galaxies have had a longer star formation history: we propose instead that these objects formed or appeared over a longer timescale in the Universe's history and the older ones have not changed much, apart from the ageing of stars. Larger galaxies on average, necessarily took more time to assemble and complexify (diversify), so that it is very unlikely to find young and very diversified galaxies. However, the notion of galaxy age must be questioned.

Figure~\ref{figellipses} illustrates the fundamental difference between diversification and age. The Clad1 group, which is assumed to be ancestral because of its low metallicity,
appears to be the youngest group on average according to stellar evolution models. If a galaxy is to resemble the most pristine objects, it must not have been transformed too much even by secular evolution. This is why the most pristine objects are necessarily relatively young and metal-poor. Conversely, the most diversified objects have a higher average stellar age, which provides no information on the epoch of the transforming events that gave them their observed properties.  As a consequence, old galaxies are not obvious ancestors.  In addition, the spread in age within each group is generally large and overlaps with the spread in age of the other groups. Age is thus not a good landmark of evolution. From the point of view of astrocladistics, the so-called downsizing effect results from a confusion between age and the level of diversification \citep{jc1,jc2,FDC09}.

Hence, the age of a galaxy or a group of galaxies is probably not so important and may even be meaningless \citep{Serra2007}. We even find this term misleading, and suggest it be replaced by ``average stellar age''. The diversification state should be used instead, as it reflects the actual assembly history of a galaxy.

\subsection{Correlations within groups (``specific correlations'')}
\label{corrcausal}

As previously seen (Sect.~\ref{corrspur} and Fig.~\ref{figspurious}), the two scatter plots of \ldiam\ vs \Kabs\ and \Mgb\ vs \MgbFe\ show both a global evolutionary correlation and correlations within the groups (which we call ``specific correlations''). Four other scatter plots only show specific correlations, the global correlation being less obvious and/or more dispersed : \Brie\ and \Kabs\ vs \logre, $\log(M_{dyn})$ vs \logre, and \MgbFe\ vs $Fe5270$. These six specific correlations are shown in Fig.~\ref{figsgroupdep}.

Diversification within each group is determined by the structure of the tree in Fig.~\ref{figtreeclad}. If we examine the evolution of the parameters involved in the correlations along each branch (thus each group) of the tree, we find that:
\begin{itemize}
   \item \Kabs\ increases slightly with diversification within Clad6 and Clad8;
   \item \MgbFe\ might possibly increase in Clad8;
   \item \logre\ might possibly decrease in Clad6;
   \item $M_{dyn}$ decreases slightly in Clad6 and might possibly increase in Clad5 and Clad8.
\end{itemize}

This is clearly not enough to explain all the observed specific correlations with evolution within the groups. However, the difference between objects of a same group is weaker than for the whole sample and thus would require refined cladistic analyses with possibly additional descriptors. We now examine in some detail the scatter plots in Fig.~\ref{figsgroupdep}. 

The correlation, which is particularly tight and linear, between \Mgb\ and \MgbFe, also holds within each group. Since the first parameter largely depends on the $\alpha$-elements, while the second is essentially independent of it \citep{Thomas2010}, these specific correlations can probably be explained, in a similar way to the global one, by evolution. The correlations between \MgbFe\ and either $Fe5270$ or $Fe5335$ have larger scatters.

The correlation between \ldiam\ and \Kabs\ seems to be present within each group. This is not proof however that it is a causal relation because the larger the galaxy the more luminous, it still be due to evolution within each group or to some other confounding parameter \citep[][]{DFB2011}. In addition, all correlations, either global or specific, are approximately similar, which suggests that they have the same explanation.

The Kormendy relation (\Brie\ vs \logre) clearly appears to depend on the group, and has a far smaller scatter for the most diversified groups. There is an ``evolution'' in the correlation curve following diversification, the different relations appearing stacked on each other. At first glance, galaxies are brighter when more diversified, but this is not so simple if we look at Clad7 and Clad8: galaxies from the first group are globally larger and fainter. The correlation also has a larger scatter for Clad1 and Clad3.

The \Kabs\ vs \logre\ relation is quite dispersed, but there are slightly more convincing correlations for some groups, at least for the most diversified ones. For Clad1, there is little variation in \logre, so that there is no real correlation. The difference between this relation and the \Kabs\ 
vs \ldiam\ one is striking.

The $M_{dyn}$ vs \logre\ relation is tight, with very clear correlations specific to each group and with relatively little overlap between them. This plot can be usefully compared to numerical simulations \citep[e.g.][]{Robertson2006} as done in \citet[][see Sect.~\ref{fundplane} and Sect.~\ref{history}]{Fraix2010}.

The \MgbFe\ vs $Fe5270$ relation is quite dispersed, despite the square-root relation linking both parameters. Correlations can be easily seen within groups, and they appear to generally differ (particularly in terms of the slope) from the global relation.

To summarize, there are several cases where the specific correlations are present in all or some of the groups, regardless of whether the global correlation exists. Why is this so?

If the correlation is present both globally and within groups, then we can guess that it is for the same reason. In the two cases here (\ldiam\ vs \Kabs\ and \Mgb\ vs \MgbFe), the global correlation is evolutionary (Sect.~\ref{corrspur}) and since all correlations appear to have approximately the same slope, then the specific correlations should be evolutionary as well.

In the other cases where there is no obvious global correlation, the reason must be specific to the group, and probably different from one group to another. The correlations might be explained by a direct physical cause, or by a confounding parameter, which can still be evolution. Note that the confounding factor may depend on the group. 

Anyhow, the origin of the correlations and their properties is quite complex. In the case of the $M_{dyn}$ vs \logre\ relation, numerical simulations show that it is determined by several variables involved in the assembly history, such as the epoch of the last merger, the level of dissipation, the number of accretion events, the impact parameters, and so forth \citep[][]{Robertson2006}. Thanks to a good diversity of simulated galaxy populations, \citet{Fraix2010} were able to derive the history assembly of each group. The specific correlations can then be explained by either several drivers from the physical point of view, ``cosmic variance'' within each group from the observational point of view, or confounding factors from a statistical point of view.

Consequently, the two scatter plots showing both global and specific correlations are probably driven by a dominant general evolutionary factor (such as perhaps dynamical evolution for \ldiam\ vs \Kabs\ and chemical evolution for \Mgb\ vs \MgbFe) affecting all galaxies of the sample, while the other ones have multiple and necessarily specific factors, as in the $M_{dyn}$ vs \logre\ relation. For instance, the importance of merger events applies only to these galaxies that have experienced such a catastrophic transforming process during their assembly history.

\subsection{Fundamental planes}
\label{threedcorr}

\subsubsection{The fundamental plane of early-type galaxies}

\label{fundplane}

The well-known and intensively studied correlation between \Brie, \logs, and \logre\ is called the fundamental plane. The first multivariate analysis of this relation was performed recently by \citet{Fraix2010}. The present sample and that of \citet{Fraix2010}, which are both at low redshift, have no galaxy in common and the parameters used to partition the sample into groups are different.

The present partitioning is in excellent agreement with the result of \citet{Fraix2010} as illustrated in Fig.~\ref{figfundplane}, which shows
the projection onto the fundamental plane (\logs\ vs \Brie) of the partitionings obtained by cladistics and cluster analysis in the present paper, and the partitioning obtained by \citet{Fraix2010}. The structures within the fundamental plane found in \citet{Fraix2010} are thus confirmed.

There is a good correspondence between the groups in the two studies, as can be seen in Fig.~\ref{figfundplane} and in more detail in Fig.~\ref{figcompgroupsFP}:  C1 includes Clad4 and a large part of Clad3, C4 $\simeq$ Clad6, C3, C5, and C6 are essentially included into Clad7, and C7 $\simeq$ Clad8. The \logre\ vs $M_{dyn}$\ diagram (Fig.~\ref{figcompMdynRe} and Fig.~\ref{figcompgroupslogreMdyn}) confirms these equivalences, pointing out that C1 overlaps both Clad3 and Clad4 and is distributed in a way that is more similar to Clad3.
There are however some differences. 

Clad1 seems to occupy a region of the fundamental plane (\logs\ vs \Brie) that is not very well-covered by the sample used by \citet{Fraix2010}. 
  
Clad2 has no equivalent in \citet{Fraix2010} when projected onto the fundamental plane (Fig.~\ref{figfundplane}). This group is also plotted separately in Fig.~\ref{figcompMdynRe} to show that it follows the same correlation as the other groups, spanning nearly the full range of both \logre\ and $M_{dyn}$.

Clad5 is also absent in \citet{Fraix2010}. Since it appears in the very centre of both Fig.~\ref{figfundplane} and Fig.~\ref{figcompMdynRe}, we believe that it is identified here because of the larger number of parameters used here, which in effect provides a higher-resolution analysis. Its properties were also found to be quite similar to Clad4 (Sect.~\ref{discussion}), so that Clad4 is quite different from C1 (see above).

Group C2 of \citet{Fraix2010} is absent in the present partitioning. This is perhaps because the corresponding regions of the fundamental plane (Fig.~\ref{figfundplane}) are not very populated in our sample, or because of the different sets of parameters used in the two studies.

To complete the comparison between the two studies, we performed the same analysis as in \citet{Fraix2010}, using the same four parameters \logs, \logre, \Brie, and $Mg_2$ (Appendix~\ref{fourparam}). 
Naturally, since these four parameters are not all discriminant for the present sample, the resulting tree and the corresponding partitioning are slightly less robust. However, the agreement is still quite good. The sample used in the present paper \citep{Ogando2008} is globally at a lower redshift than the one used in \citet{Fraix2010} \citep{hudson2001}. This renders the determination of the distance, and thus \logre\ less accurate \citep[see][]{Ogando2008}. This could partly explain why \logre\ has not been found to be discriminant or why the four-parameter result here is slightly less robust than in \citet{Fraix2010}. 

We thus confirm the result of \citet{Fraix2010} that there are structures within the fundamental plane. Since the organisation of the groups on this plane defines very similar evolutionary paths corresponding to a clear trend in diversification as defined by Fig.~\ref{figtreeclad}, we confirm that this global relation in three-parameter space is mainly driven by diversification. For the sake of clarity, we stress that this evolutionary interpretation of the fundamental plane holds for the groups, not within the groups, since neither the present paper nor that by \citet{Fraix2010} tackle this question. This would certainly merit further specific studies because \citet{Fraix2010} found that the tightness of the fundamental plane strongly depends on the group considered, the least diversified ones showing a very loose -- if significant at all -- correlation.

\subsubsection{Other correlation planes}
\label{otherplanes}

%%%%%%%%%%%
   \begin{figure}
   \centering
  \includegraphics[width=\columnwidth]{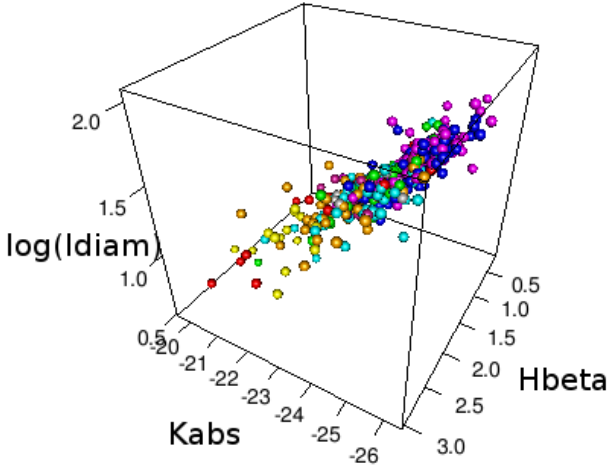}
   \caption{Another ``fundamental plane'' in the parameter space defined by  \Hbeta, \Kabs, and \ldiam.} 
    \label{figotherplanes}%
    \end{figure}
%%%%%%%%%%%%%

Concerning the fundamental plane, the groups follow one another along a path of diversification in the \logre\ vs \logs\ (edge-on projection) scatter plot, whereas they are clearly distinguishable and distributed in no obvious order in the \Brie vs \logre\ (face-on projection) scatter plot .

Hence, one can expect to find other ``fundamental planes'' by looking at the behaviour of groups in two scatter plots made with a set of three parameters. This is the case for \Hbeta, \Kabs, and \ldiam\ (Fig.~\ref{figotherplanes}). The two latter parameters are obviously reminiscent of \Brie\ and \logre, while \Hbeta\ is of a different nature from \logs, so that this fundamental plane is not redundant with the classical one. Replacing \Hbeta\ by \DsB, one also obtains a nice correlation, thus another fundamental plane.

It is interesting to note that the parameters of these fundamental planes might be easier to observe than for the classical one, but still allow the distance determination thanks to \ldiam.

There may be additional similar surfaces in higher-dimension parameter spaces but that have more complex projections. Would these be interesting to discover and would they be more useful than the classical fundamental plane?

The answer to these questions is twofold. First, they are identical to the classical fundamental plane, in the sense that they are essentially evolutionary correlations driven by diversification \citep{DFB2011}. All are simply projections of the tree in a sub-parameter space. They are thus not more and not less informative. Second, they can be useful once the confounding factor (here evolution) has been taken into account. This is however beyond the scope of the present paper.

\section{The assembly history of early-type galaxies}
\label{history}

%%%%%%%%%%%
   \begin{figure*}
   \centering
 \includegraphics[width=15 true cm]{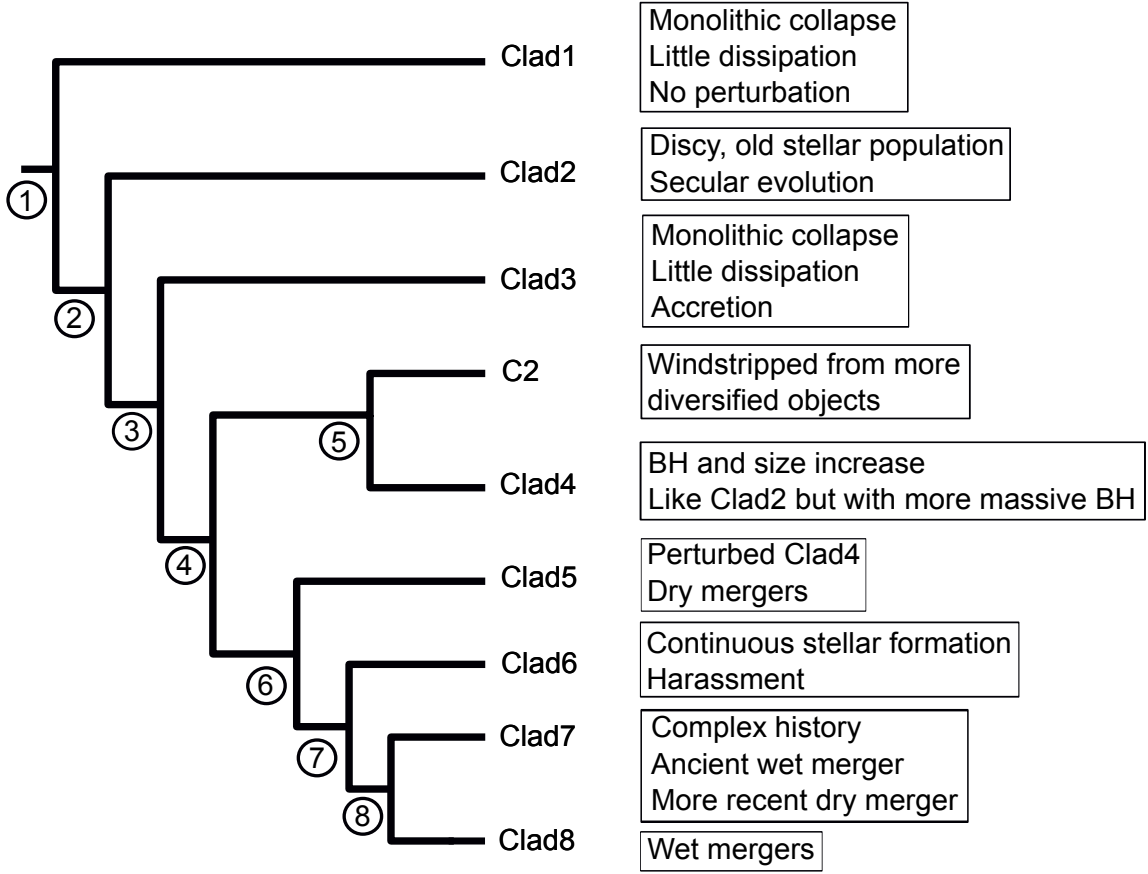}
   \caption{Combination of the tree from this paper (Fig.~\ref{figtreeclad}) and the one in \citet{Fraix2010}, showing the proposed assembly history. Numbers indicated at nodes are referred to in the text.} 
    \label{figsglobalclassif}%
    \end{figure*}
%%%%%%%%%%%%%

\citet{Fraix2010} uncovered the assembly history of a completely different sample of early-type galaxies by analysing its properties in both the fundamental plane and a mass-radius diagram with the help of numerical simulations from the literature. This sample was composed of galaxies in clusters, while the present one is composed of galaxies in the field, groups, or clusters, and has the advantage of more properties being documented.  Emission-line galaxies were excluded from both samples. We repeat here the exercise for both sets of groups merged together following the correspondence detailed in Sect.~\ref{fundplane}.

\begin{itemize}
   \item Clad1: these galaxies have low metallicity and in many respects look quite primitive, having a very low dynamical mass. They are less metallic and more primitive than the Clad3/C1 ones, which were chosen as the most primeval group in \citet{Fraix2010}. Clad1 galaxies could be the remains of a simple assembly through a monolithic collapse with little dissipation and their somewhat discy nature probably requires significant feedback and a few perturbations \citep{Benson2010}.
   \item Clad2: this group shows a steep correlation between \logre\ and $M_{dyn}$. This could be indicative of some merger processes \citep[e.g.][]{Ciotti2007} but the galaxies are discy. However, they also have a high \logs. This suggests that these galaxies are quite primitive objects similar to those of Clad1 but since they are more massive, they underwent a more significant secular evolution, perhaps as in the case of ``pseudo-bulges'' \citep{Kormendy2004,Benson2010}.
   \item Clad3/C1:  chosen as the most primeval group because of its low average $Mg_2$ in \citet{Fraix2010}, this group is here surpassed by the still lower $Mg_2$ of Clad1. They found that galaxies of C1 are possibly the remains of a simple assembly through a monolithic collapse with little dissipation, and were probably perturbed by interactions. We propose instead that accretion is the main perturbations because the Clad3 galaxies are small, not very much concentrated, and have a low \logs.
   \item C2: they were found in \citet{Fraix2010} to be less massive and smaller than the ones of Clad3/C1, and have a slightly higher $Mg_2$. They are also somewhat brighter. They could be the remains of wind stripping of some kind of more diversified objects because of a strong interaction.
   \item Clad4: this group is very similar to Clad2, but since the Clad4 galaxies are larger and have a higher $M_{dyn}$, they might have been perturbed by a strong interaction that yielded a more massive central black hole.
   \item Clad5: being very similar to Clad4 objects, they could be these discy galaxies seen edge-on but this is statistically untenable because there are three times more Clad5 objects than Clad4 ones. Since they have a very low \DsB, a more likely explanation is that Clad5 galaxies could be perturbed versions of Clad4 members. These perturbations are probably mergers since these objects have lost the disciness of Clad4 objects, but to preserve similar properties, little gas should be involved, which implies that dry mergers are the most probable transforming events.
   \item Clad6/C4: three scenarios were proposed for C4 in \citet{Fraix2010}: these objects could simply be either galaxies in which star formation has been continuous, C1 galaxies in which the initially richer gas has not been swept out, or the remnants of several minor mergers and accretion. The Clad6/C4 galaxies have unexpectedly low values of $M_{dyn}$, \Brie, and \logre, as well as to a lesser extent \ldiam. Otherwise, they do not look odd, so that a continuous star formation with few external perturbations could also be a reasonable explanation. However, we find that they are very similar to Clad2, which we proposed to have undergone significant secular evolution, but have a much lower \DsB. This suggests that many interactions, such as harassment \citep{Moore1996}, could be the culprit.
   \item Clad7: because Clad7 includes the groups C3, C5, and C6, their history might be complex according to \citet{Fraix2010}, involving many transforming events (accretions, minor mergers, together with more or less dissipational major mergers). They are probably the remains of both wet and dry mergers, the most recent ones being of the latter kind. The low value of \Hbeta, that would indicate that the last star formation event is relatively ancient, reinforces this interpretation. They could represent a kind of end state of galaxy diversification. 
   \item Clad8/C7: C7 galaxies were found to be small and very metallic with a high surface brightness, and to define a tight FP. They seemed to be associated with the remains of a dissipative (wet) merger, with very little or no dry mergers. They might also have formed through minor mergers and accretions but the tight FP favors the dissipative wet-merger scenario. The low \DsB\ found here for Clad8/C7 tends to confirm this conclusion. We believe that they may well define another possible end state for galaxy diversification.
\end{itemize}

We summarize the above histories in a single cladogram (Fig.~\ref{figsglobalclassif}), combining the trees obtained in the two studies.

The best way to interpret the evolutionary scenario depicted in this cladogram is to identify, for each node of the tree, a particular transforming event that could characterize all groups related by branches and sub-branches starting from this node \citep{jc2}. The sequence of nodes downwards starting at the upper left of the tree thus defines a sequence of ``innovations'' that occurred in a common ancestor and were transmitted to all its descendant species. In principle, these innovations are the properties of the galaxies that remain as imprints transmitted through subsequent transforming processes. Here, we consider the transforming events as innovations since they are the origin of the modifications of the properties of galaxies. However, it should be kept in mind that parallel evolutions (events occurring independently on two different lineages), convergences (different pathways leading to the same parameter value), and reversals (backward parameter evolution) are probably present, making this 
exercise currently quite tentative. These behaviours (called homoplasies) are supposedly not too numerous here since the parcimony optimisation of the cladistic analysis minimizes the occurrence of these types of parameter evolutions. We do not discuss them in this very first attempt, especially because the properties at hand are few.

We attempt to identify these innovations in Fig.~\ref{figsglobalclassif} using the possible histories of each group previously identified. They appear on the tree when they first occur in the history of the Universe. It is thus expected that the most basic transforming events occur very early and the more complex ones later, making the sequence of ``innovations'' along the tree a representation of the so-called cosmic evolution. 

Finally, we remind the reader that if the level of diversity goes along the vertical axis in Fig.~\ref{figsglobalclassif}, the horizontal axis, especially the branch lengths, has no particular meaning in this representation. It is important to keep in mind that we are dealing with present-day galaxies and properties, not those that prevailed at the time of the transforming event indicated at the node from which a given branch emerges. Simply speaking, this means that the galaxies of, say, Clad1, are present-day galaxies that have passively evolved from a less diversified initial state than those of either Clad6 or Clad7. 

The nodes are identified by the numbers in Fig.~\ref{figsglobalclassif}. The corresponding proposed events are:
\begin{enumerate}
   \item collapse;
   \item secular evolution;
   \item accretion;
   \item interaction;
   \item ``gaseous'' interaction;
   \item dry merger;
   \item harassment;
   \item wet mergers.
\end{enumerate}

The first transforming event (node 1), which marked the history of all the galaxies, is most likely monolithic collapse. This is probably the simplest process to form a self-gravitating ensemble of gas, stars, and dust that we can call a ``galaxy''.  The galaxies of Clad1 evolved passively after exhausting their gas reservoir. 

The next three events (nodes 2-4) must have been gentle, since the discy morphology is well-preserved until Clad4 and it is now well-established that minor mergers generally preserve the structure of discy galaxies, while mergers of galaxies of comparable masses generally do not. We first have (node 2) the secular evolution, which is defined as the evolution of a galaxy in isolation, which is expected to be far more frequent and capable of significantly modifying the structure and properties of galaxies. At node 3, we must then invoke accretion must be invoked to increase the masses of galaxies. A more complex event follows, namely interaction, which is an external perturbation that, with the wealth of possible impact parameters and galaxy properties, is probably the main driver of galaxy diversity, especially during the first Gyr of the Universe \citep{Benson2010}.

For node 5 between C2 and Clad4, interaction involving gas must be invoked to either strip the gas in C2 galaxies by ram pressure or feed the central black hole in Clad4 objects. 

Mergers must be advocated at node 6 since the more diversified groups have lost their disciness. For Clad5 galaxies, the most probable transforming event is a major merger without much gas (dry mergers).

A substantial star formation must have occurred in Clad6 galaxies, and several properties indicate repeated perturbations, implying that harassment is a good candidate for node 7 \citep{Moore1996}. Harassment is the cumulative effect of high-speed galaxy encounters, that heats the disc (log$\sigma$ increases) and favors gas inflow to the galaxy center. This kind of transforming event acts on a longer timescale than the dry mergers at node 6, which explains why it appears ``later'' in the cladogram. 

The two most diversified groups, Clad7 and Clad8, are found to have had complex histories, certainly including wet mergers (node 8).

Many associated processes, such as feedback and the quenching of star formation \citep[e.g.][]{Bundy2006,Benson2010} are not proposed here because we have concentrated on more generic events. A significant difficulty of such an exercise is to identify some properties with transforming events that are very complex, involving diverse impact parameters and various chemical, physical, and dynamical processes. We believe that it is somewhat illusory to associate a particular feature with any each of these events, and only statistical analyses of simulated cases could provide average properties that could be compared to statistical analyses of real objects similar to the one we have performed here.

\section{Conclusion}
\label{conclusion}

We have used several multivariate tools, first to select the most discriminant parameters from the 25 initially available for the sample of 424 fully documented galaxies of \citet{Ogando2008}, and second to partition the sample into groups. The three partitioning methods yield similar numbers of groups and similar composition for each of them, considering that some fuzziness is expected.

Our first result is that among the initial 23 quantitative parameters available in this study, only 6 are discriminant and actually yield a relatively robust partitioning for this sample. Among the 10 Lick indices, only 2 (\MgbFe\ and \NaD) are discriminant, together with \logs, \Brie, \OIII, and \DsB.

The global evolutionary scenario found by astrocladistics gives a very sensible result: galaxies tend to globally become more metallic, more luminous (more massive), and larger with increasing diversification. At the same time, they acquire a larger central velocity dispersion, which is often related to the mass increase, and \NaD\ also increases along with both mass and velocity dispersion, as expected. These are global statistical trends that are explained by general basic physical and chemical processes as a function of time since the Big Bang. 

As a consequence, the many properties of galaxies that are bound to evolve on average with galaxy diversification explain the several evolutionary correlations found in this paper. In particular, we have confirmed the evolutionary nature of the $Mg_2$ vs \logs\  correlation found by \citet{Fraix2010} for a different sample and using a different set of parameters. Rather interestingly, we have also found some correlations that are specific only to some groups. These can be attributed to either a direct physical cause or a confounding factor specific to some groups (such as the epoch of the last merger, the level of dissipation, the number of accretion events, the impact parameters, or the number of mergers).

One of the most important results of our work is that the structures defined by the partitioning, when projected onto either the fundamental plane (\logs\ vs \Brie) or the \logre\ vs $M_{dyn}$\ diagram, are very similar to those found by \citet{Fraix2010} for a totally distinct sample with different parameters.

The fundamental plane of early-type galaxies appears to be very probably generated by diversification. In support of this, we have also found another ``fundamental plane'', a three-dimensional correlation between \Hbeta, \Kabs, and \ldiam. 
All scatter plots are basically simple projections onto a sub-parameter space of the partitioning established in the six-parameter space. Thus, there is less information in either these scatter plots or ``fundamental planes'' than in the multivariate partitioning and the evolutionary tree obtained with cladistics.

Another important result is that six parameters -- no fewer -- are needed to describe the diversity of this sample. The three parameters of the fundamental plane (\logs, \Brie, and \logre), plus the index \Mgdeux, have not yielded as robust a partitioning here, although they did in a previous study for a distinct sample \citep{Fraix2010}. We argue that there is no contradiction here, because three discriminant parameters extracted in the present paper are present in our previous study (\logs, \Brie, $Mg_2$ being replaced by \MgbFe). The multivariate analyses generally depend on both the objects in the sample and their initial set of descriptors, both of which are different in the two studies. Nevertheless, as a consequence, similar analyses will have to be conducted on other samples with other descriptors, since our fairly small sample of nearby galaxies cannot represent the diversity of galaxies throughout the Universe, and the available parameters are here restricted to the visible domain.

We have combined the results of \citet{Fraix2010} and those in the present paper on a single cladogram, showing the possible assembly history for each group. From this cladogram, we have attempted to identify the transforming events that are at the origin of galaxy diversification.
The transforming events that we have indicated as ``innovations'' are tentative, because the information at hand is insufficient to identify them with certainty. These proposed events show that the use of sophisticated statistical tools yields a very sensible classification. Figure~\ref{figsglobalclassif} is the basis of an \emph{explanatory} classification linking the objects to the fundamental transforming processes, i.e. to the physics, rather than a \emph{descriptive} classification adopted in most current classifications of galaxies. In this respect, we note that the Edwin Hubble classification is of the latter type, being based on morphology, while his tuning fork diagram (often called the Hubble sequence nowadays) is explanatory since it indicates the links between the classes. Nearly one century later, we know that galaxies are characterized by much more than only their morphology, so that we need to generalize the Hubble diagram to a multivariate picture of galaxy diversification. Figure~\ref{
figsglobalclassif}, which we have produced using cladistics, is one step in this direction.

Hence, increasing both the sample size and the number of descriptors is an absolute requirement. The six-parameter space needed to describe the diversity of the sample used in the present paper is probably a minimum space because of the complexity of galaxies and their assembly history. The nature of the discriminant parameters might also change with the input of more observables. In addition, the number of groups and their boundaries will certainly change. This is a double quest: classifying galaxies into objectively established evolutionary and intelligible groups, and finding the parameter space in which these groups can be identified. This quest is necessarily progressive, and will probably never end. However, one can hope that some convergence will be reached.

A limitation of the present work is that cladistics cannot be applied directly to very large samples as the necessary computer time would be prohibitively excessive. However, once the most discriminant parameters are identified, it will be possible to repeat the cladistic analysis for many subsamples, and subsequently combine the trees to define classes of galaxies. The ultimate goal is to gather the huge number of galaxies in the Universe into a tractable number of groups and establish the corresponding evolutionary relationships.

\begin{acknowledgements}
This research has made use of the  HyperLeda database (http://leda.univ-lyon1.fr) and the NASA/IPAC Extragalactic Database (NED) which is operated by the Jet Propulsion Laboratory, California Institute of Technology, under contract with the National Aeronautics and Space Administration.
\end{acknowledgements}

\appendix

\section[]{Methods}
\label{appendMeth}

\subsection{Principal component analysis}
\label{metPCA}

Principal component analysis (PCA) is  well-known to astronomers. It is not a partitioning method: its aim is instead to reduce the dimensionality of the parameter space. From the correlation matrix, PCA builds eigenvectors (the principal components) that are orthogonal and linear combinations of the physical parameters. These eigenvectors usually have no physical meaning. In general, most of the variance of the sample can be represented with only a few principal components (those having an eigenvalue greater than 1). They thus give a simpler representation of the data by eliminating the correlations between physical parameters. Strongly correlated parameters are gathered in the same eigenvector, and the most important parameters (with respect to variance) are the ones with the highest coefficient (loading) in each eigenvector. The physical interpretation must be made back in the real parameter space.

PCA is thus very efficient at reducing the parameter space to supposedly uncorrelated components and helps in detecting the most discriminant or discriminating parameters. The number of significant eigenvectors gives an idea of the number of parameters necessary to describe the sample. Principal components can also be used for subsequent cluster or cladistic analyses.

There is however a caveat to be kept in mind. PCA eliminates all correlations, regardless of whether they are causal. It is extremely useful to remove any redundancies, as well as physical correlations between two parameters indicating the same underlying process. However, PCA also removes evolutionary correlations \citep[which are called ``spurious'' or confounding in statistics,][]{DFB2011}, for instance between two parameters that are independent but vary with time. The \logs-$Mg_2$ correlation for early-type galaxies \citep[see][]{Fraix2010} is a good example. Such independent evolutions are lost through the PCA reduction of dimensionality.

\begin{table*}
 \centering
  \caption{Loadings on the eight principal components of the PCA analysis made on the set of 23 parameters (see Sect.~\ref{parPCA}) .}
     \label{tab_loadings}
\begin{tabular}{lllllllll}
      \hline
   &   Comp1 &     Comp2 &   Comp3 &   Comp4 &   Comp5 &    Comp6 &    Comp7 &    Comp8   \\
                  \hline
Mgb   & -0.9143 & -0.122395 &  0.1594 & -0.1593 & -0.1215 &  0.04717 &  0.11541 &  0.03549   \\
logs  & -0.9067 & -0.012224 &  0.0674 &  0.0950 & -0.0443 &  0.02837 & -0.05006 &  0.04723   \\
Mg2   & -0.8983 & -0.158860 &  0.1190 & -0.1121 & -0.1455 &  0.05519 &  0.07249 & -0.00744   \\
mgbfe & -0.8879 & -0.303358 &  0.0656 & -0.1870 &  0.0543 & -0.04251 &  0.00546 & -0.04366   \\
NaD   & -0.8659 & -0.099002 &  0.0527 &  0.0657 & -0.0496 & -0.00899 & -0.00360 & -0.03957   \\
Mg1   & -0.8470 & -0.132117 &  0.1739 & -0.0977 & -0.1787 &  0.01250 &  0.06330 & -0.05674   \\
Kabs  &  0.7780 & -0.357711 &  0.1878 & -0.3577 & -0.1884 &  0.00503 &  0.04219 & -0.10201   \\
mabs  &  0.7255 & -0.359334 &  0.2095 & -0.4395 & -0.2075 &  0.06904 & -0.06490 & -0.11826   \\
ldiam & -0.7248 &  0.409603 & -0.2557 &  0.3327 &  0.1887 & -0.09360 &  0.02929 &  0.06032   \\
mgb.fe& -0.6742 &  0.245147 &  0.2938 & -0.0296 & -0.3977 &  0.17776 &  0.26559 &  0.15522   \\
logre & -0.6512 &  0.570850 & -0.3297 & -0.2541 & -0.0686 & -0.01075 & -0.01336 & -0.01325   \\
Fe5335& -0.5455 & -0.406391 & -0.1312 & -0.2574 &  0.3127 & -0.11438 & -0.13735 & -0.09773   \\
hbeta &  0.5389 & -0.223947 & -0.4974 &  0.2296 &  0.0104 &  0.18101 & -0.01882 & -0.00539   \\
Fe5406& -0.4906 & -0.465277 & -0.1337 & -0.0838 &  0.1044 & -0.13418 &  0.01741 & -0.11408   \\
Fe5270& -0.4900 & -0.522075 & -0.1240 & -0.1365 &  0.2660 & -0.18686 & -0.15191 & -0.15975   \\
H.K   & -0.3065 &  0.117644 &  0.0945 &  0.2266 &  0.4477 &  0.48040 & -0.20626 & -0.28296   \\
Fe5015& -0.2763 & -0.550090 & -0.5230 &  0.1094 & -0.2198 &  0.17362 &  0.05076 & -0.00969   \\
D.B   &  0.2024 &  0.134962 & -0.3041 & -0.1834 &  0.2342 & -0.15225 &  0.66215 & -0.38279   \\
B.R   & -0.1338 &  0.363115 & -0.2925 & -0.4631 & -0.1578 &  0.13444 & -0.51784 & -0.06306   \\
Brie  & -0.0673 &  0.614240 & -0.4327 & -0.4875 & -0.0751 & -0.08047 &  0.03262 & -0.07633   \\
Fe5709&  0.0602 & -0.172935 & -0.1399 & -0.3042 &  0.4027 & -0.14195 &  0.06450 &  0.74089   \\
OIII  & -0.0311 & -0.395042 & -0.6303 &  0.1494 & -0.3828 &  0.25483 &  0.08224 &  0.13941   \\
J.H   & -0.0175 & -0.000856 & -0.0977 &  0.3589 & -0.3548 & -0.69952 & -0.23475 & -0.11213   \\
                \hline
\end{tabular}
\end{table*}   

\subsection{Minimum contradiction analysis}
\label{metMCA}

Partitioning objects consists in producing some order. In some cases, i.e. in either hierarchical clustering or cladistics, the arrangement of the objects can be represented on a tree. 
A tree is a graph representing the objects as the leaves with a unique path between any two vertices. A bifurcating tree has internal vertices that all have a degree of at most 3 (at most 3 branches connect to any such vertex). 

By indexing circularly all the leaves of a planar representation of a weighted binary tree, one obtains a perfect order, meaning that the corresponding ordered distance-matrix fulfills
all Kalmanson inequalities.  Generally speaking, the Kalmanson inequalities are fulfilled if the ordered distance matrix corresponds to a weighted binary tree or a superposition of binary trees
\citep{Thuillard2011}. The difference between the perfect order and the order one obtains with a given dataset is called the contradiction. The minimum contradiction corresponds to the best order one can get. 

The minimum contradiction analysis \citep[MCA,]{Thuillard2007,Thuillard2008} finds this best order. It is a powerful tool for ascertaining whether the parameters can lead to a tree-like arrangement of the objects \citep{TF09}. Using the parameters that fulfil this property, the method then performs an optimisation of the order and provides groupings with an assessment of their robustness.

For taxa indexed according to a circular order, the distance matrix , which is defined to be
\[ Y_{i,j}^n = {1 \over 2} (d_{i,n}+d_{j,n}-d_{i,j}) \]
fulfils the so-called Kalmanson inequalities (Kalmanson, 1975):

 \begin{equation}
      Y_{i,j}^n \ge Y_{i,k}^n, {Y_{k,j}}^n \ge Y_{k,i}^n \;\; (i \le j \le k) 
\end{equation}
where $d_{i,j}$ is the pairwise distance between taxon $i$ and $j$.
The matrix element $Y_{i,j}^n$
is the distance between a reference node n and the path i-j. The
diagonal elements $Y_{i,i}^n=d_{i,n}$ correspond to
the pairwise distance between the reference node $n$ and the taxon $i$.

 The contradiction on the order of the taxa can be defined as

\begin{equation}
C =  \sum_{k > j \ge i}\left( \max \left(\left(Y_{i,k}^n - Y_{i,j}^n\right),0\right)\right)^2 +\sum_{k \ge j > i}\left( \max \left(\left(Y_{i,k}^n - Y_{j,k}^n\right),0\right)\right)^2
\end{equation}
for any $i,j,k \ne n$. The best order of a distance matrix is, by definition, the order minimizing the contradiction \citep{Thuillard2007,Thuillard2008}.
\citet{TF09} showed that the perfect order is linked to the convexity of the variables in the parameter space, and is obtained for specific properties of the variables along the order. It is then possible to detect the discriminant potentiality of the variables. This is exactly what is done in Sect.~\ref{parMCA}.

\subsection{Cladistic analysis}
\label{metClad}

\begin{table*}
 \centering
 \begin{minipage}{\columnwidth}
  \caption{Fitness of parameters on the cladograms obtained for each subset as represented by the Rescaled Consistency Index (RCI).}
     \label{tab_sets2}
\begin{tabular}{lll}
      \hline
Subset &  Order from RCI & RCI \\
                  \hline
4cA   & \DsB\ \logs\ \MgbFe\ \NaD &        0.102 0.086 0.086 0.075 \\
5c    & \DsB\ \logs\ \NaD\ \MgbFe\ \Brie &  0.077 0.063 0.063 0.059 0.051 \\
5cA  & \MgbFe\ \Mgb\ \logs\ \DsB\ \NaD &  0.098 0.090 0.080 0.075 0.066 \\
6c    & \logs\ \DsB\ \NaD\ \MgbFe\ \Brie\ \OIII &    0.059 0.055 0.055 0.050 0.040 0.039 \\
6cA  & \MgbFe\ \Mgb\ \logs\ \DsB\ \NaD\ \Brie &     0.076 0.073 0.061 0.060 0.054 0.043 \\
7c    & \DsB\ \logs\ \OIII\ \NaD\ \MgbFe\ $Fe5015$\ \Brie &  0.053 0.051 0.044 0.041 0.040 0.037 0.031 \\
8c    & \Mgb\ \MgbFe\ \logs\ \NaD\ \DsB\ \OIII\ $Fe5015$ &  0.055 0.052 0.050 0.044 0.041 0.038 0.033  \\
        &                                   \hfill \Brie & \hfill 0.030 \\
10c   & \MgbFe\ \Mgb\ \NaD\ \logs\ \DsB\ $Fe5270$\ \Brie & 0.075 0.055 0.044 0.042 0.034 0.033 0.025 \\
         & \hfill \BmR\ \OIII\ $Fe5709$ &        \hfill 0.025 0.023 0.020 \\
\hline
\end{tabular}
\end{minipage}
\end{table*}

Cladistics seeks to establish evolutionary relationships between objects. It is a non-parametric character-based phylogenetic method, also called a maximum parsimony method. It does not use distances, because there is no assumption about the metrics of the parameter space. The ``characters'' are instead traits, descriptors, observables, or properties, which can be assigned at least two states characterizing the evolutionary stage of the objects for that character. The use of this approach in astrophysics is known as astrocladistics \citep[for details and applications, see][]{jc1,jc2,FDC09,Fraix2010}. Simply speaking, the characters here are the parameters, the (continuous) values of which supposedly evolve with the level of diversification of the objects. The maximum parsimony algorithm looks for the simplest arrangement of objects on a bifurcating tree. The complexity of the arrangement is measured by the total number of ``steps'' (i.e. changes in all parameter values) along the tree.

The success of a cladistic analysis much depends on the behaviour of the parameters. In particular, it is sensitive to redundancies, incompatibilities, too much variability (reversals), and parallel and convergent evolutions. It is thus a very good tool for investigating whether a given set of parameters can lead to a robust and pertinent diversification scenario. 

In the present study, we used the same kind of analysis as in our previous papers on astrocladistics.
We discretized the parameters into 30 equal-width bins, which play the role of discrete evolutionary states. This choice of 30 bins is justified by a fair representation of diversity, a stability of the analysis in the sense that the result does not depend on the number of bins, and a bin width roughly corresponding to the typical order of magnitude of the uncertainties \citep[i.e. 7\%, see][]{FDC09}. We also adopted the parsimony criterion, which consists in finding the simplest evolutionary scenario that can be represented on a tree. Our maximum parsimony searches were performed using the heuristic algorithm implemented in the
PAUP*4.0b10 \citep{paup} package, with the Multi-Batch Paup Ratchet method\footnote{http://mathbio.sas.upenn.edu/mbpr}. The results were interpreted with the help of the Mesquite software \citep{mesquite} and the R-package (used for graphics and statistical analyses).

Making cladistic analyses with different sets of parameters both helps to find the most robust result and gives interesting information on the behaviour of the parameters themselves. The robustness of
cladograms is always difficult to assess objectively, so we use a criterion similar to that of other statistical distance analyses: if a similar result is found by using different conditions or
methods, then it can be considered as reasonably robust. We applied four possible tests here:

\begin{enumerate}
 \item The occurrence of a branching pattern among most parsimonious trees: with so few parameters, many equally parsimonious trees are found, often arbitrarily limited to 1000. The majority-rule consensus
of all of them yields a percentage of occurrence for each node. The higher this percentage, the higher the probability that this node is ``robust''.
\item The agreement of branching patterns between sub-sample analyses, which can be called ``internal consistency'': by making analyses of several sets of arbitrarily selected sub-samples, we can check whether
a given pattern is present on trees found with larger samples, including the full tree. 
\item The comparison between different sets of parameters: any result should preferably not depend too much on a single parameter. Adding or removing a parameter should not drastically change the tree. 
\item A comparison with the results of a cluster analysis: distance-based methods are totally independent, so any agreement can instill us a fair confidence in the result.
\end{enumerate}

Since we have many more objects than parameters, a lot of ``flying'' objects are expected between different analyses, and the above tests should be done with statistics in mind. 
The first test is always positive in this study: percentages are higher than 70-75\%, and most often they are above 95\%. This is already an indication that some structure is present in the data. The other three tests are described below.

The full sample of 424 galaxies was divided into three subsamples with 105 objects each and a fourth one with 109 objects. The first and fourth subsamples were found to belong exclusively to
clusters 1 and 3, respectively, of the cluster analysis. The diversity in the first subsample is less than for the others, so that the resulting tree is generally less well-resolved.
The two first subsamples were also gathered to form a 210-object subsample, as well as the two last ones that form a 214-object subsample. Analyses were performed with these six subsamples, as well as the full sample.
We then estimated the internal consistency by comparing the seven trees two by two and by eye (with the help of the program cophyloplot in the R-package, which connects a given object between the two trees). 

This procedure was applied to each of the eight-parameter subsets given in Table~\ref{tab_sets1}. Subsets 5c, 5cA, 6cA, and 3c show a rather good internal consistency, 4c, 7c, and 6c that which is fairly good, and finally 8c and 10c that which is not so good. 

This already shows that the optimal number of parameters is around 5, 6, or at most 7. This is in excellent agreement with the PCA analysis (Sect.~\ref{metPCA}).

If we compare the trees obtained with the full sample for the eight-parameter subsets, we find that subset 5c is very consistent with 6c, 7c, and 8c. In addition, 5c, 6c, 5cA, and 6cA are in good mutual agreement, while this is not the case for 6c, 7c, and 8c.

In Table~\ref{tab_sets2}, we show for each tree the Rescaled Consistency Index (RCI) ,which measures the fitness of a parameter on the phylogeny depicted by the tree. The higher the RCI (indeed the closer it is to 1), the more discriminant the parameter. In other words, parameters with higher RCI are the most responsible for the structure of the tree. The absolute value depends on the number of objects and parameters, so it cannot be used to compare trees obtained with different data. Here, we can only use it to compare parameters for a given tree. In Table~\ref{tab_sets2}, the parameters are ordered according to RCI. 

When \Mgb\ and \MgbFe\ are present together in a subset, they dominate the shape of the tree (sets 5cA, 6cA, 8c, and 10c), \logs\ and \DsB\ being right after them. \Mgb\ and \MgbFe\ are
obviously redundant because they are very well-correlated and are more or less the same measure. Hence, they cannot be used simultaneously in the cladistic analysis, and the trees that we find are more
linear than the others. In contrast, \logs\ and \DsB\ are
not at all correlated, but are always together, and dominate the tree shape when  \Mgb\ is not present together with \MgbFe. In addition, \NaD\ is very discriminant, and only roughly correlated with
\logs\ and \MgbFe.

If we compare the clusters obtained with the clustering analysis, the agreement decreases roughly for 6c, 7c, 5A, 3c, 5c, 4cA, 8c, and 10c, the winner being undoubtedly 6c. The corresponding tree with the groups is shown in Fig.~\ref{figtreeclad}.

\subsection{Cluster analysis}
\label{metCA}

%%%%%%%%%%%
   \begin{figure}
   \centering
 \includegraphics[width=\columnwidth]{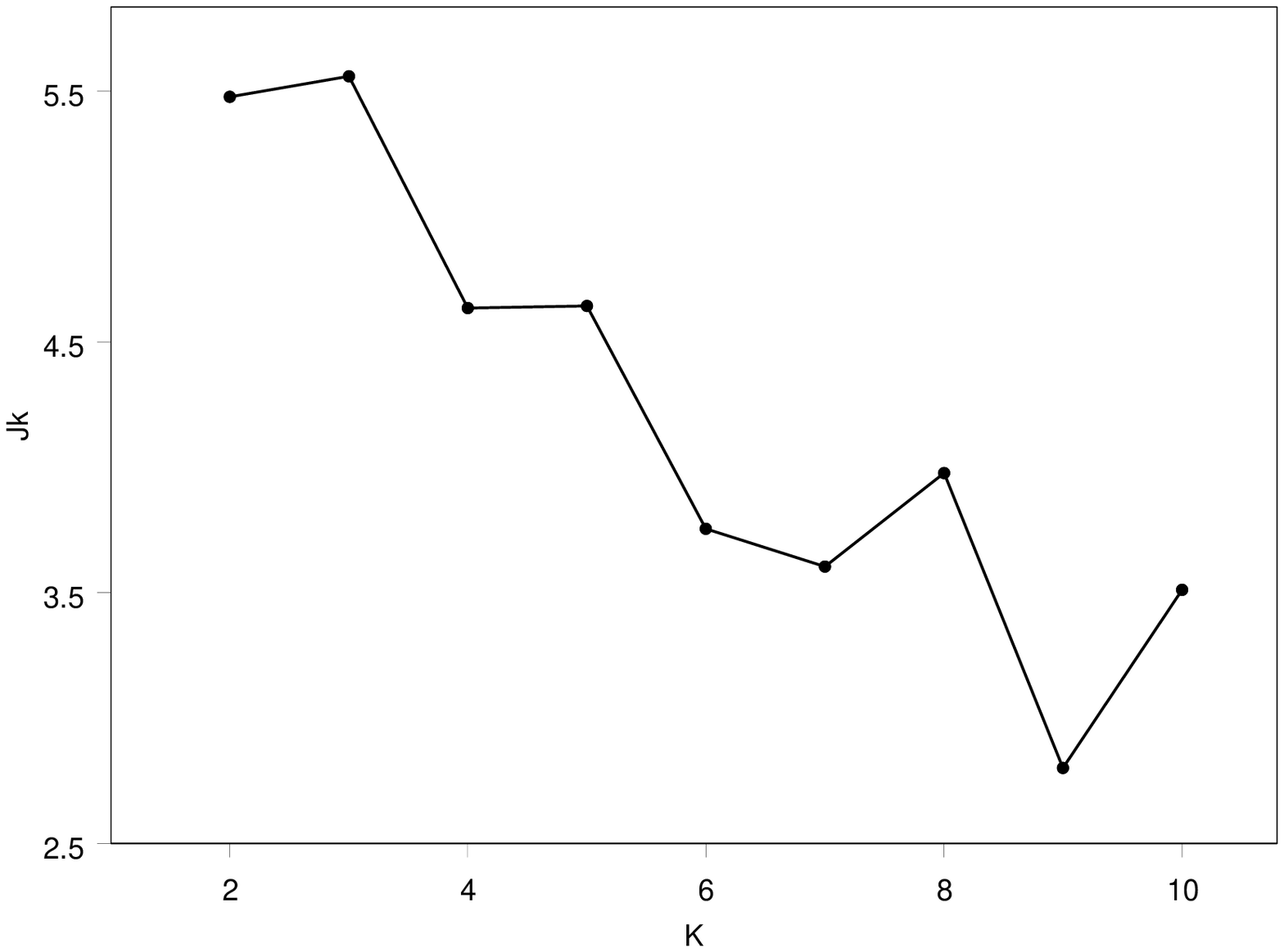}\\
\includegraphics[width=\columnwidth]{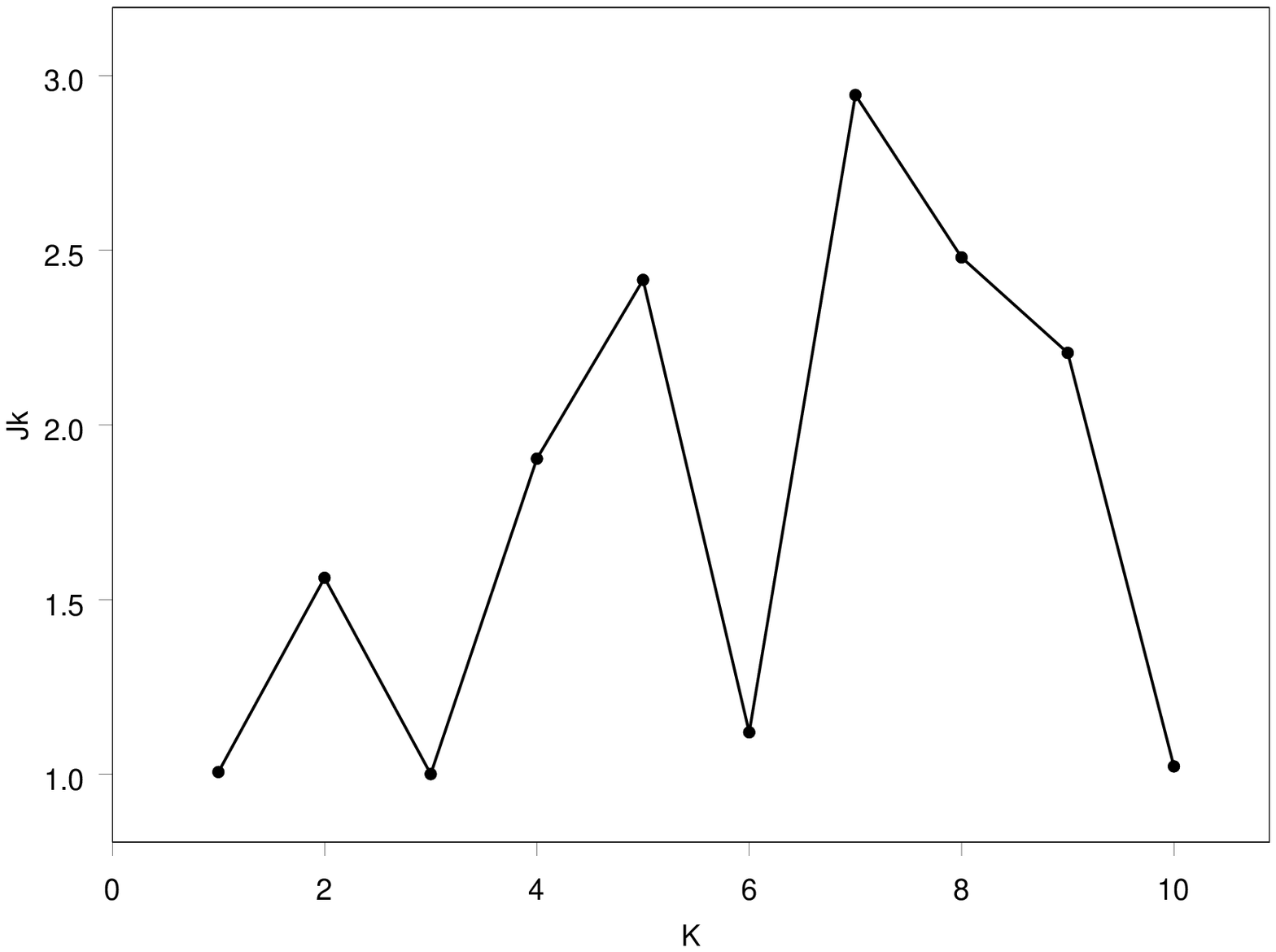} 
   \caption{Plots showing the jumps as defined in Sect.~\ref{metCA}. Top: jumps for the PCA+CA analysis (Sect.\ref{classPCAClus}). Bottom: jumps for the cluster analysis with the six parameters (Sect.\ref{classClus}).} 
    \label{figjumps}%
    \end{figure}
%%%%%%%%%%%%%

In the present study, we adopted K-means partitioning algorithm of clustering following \citet{mac67}. This method constructs
K clusters using a distance measure (here Euclidean). The  data are classified into K groups around K centres, such that the distance of a member 
object of any particular cluster (group) from its centre is minimal compared to its distance from the centres of the remaining groups. The requirement for the algorithm
is that each group must contain at least one object and each object must belong to exactly one group, so there are at most as many groups as there are objects. Partitioning methods are applied \citep{whi84,mur87,Chattopadhyay2006,Chattopadhyay2007,bab09,Chattopadhyay2009a,cha10}, if one wishes to classify the objects into K clusters where K is fixed. Cluster centres were chosen based on a group average method, which ensures that the process is almost robust \citep{mil80}. 

To achieve an optimum choice of K, the algorithm is run for K = 2, 3, 4, etc. For each value of K, the value of a distance measure $d_K$ (called the distortion)
is computed as $d_{K} = (1/p)\ min_{x} E[( x_{K} - c_{K} )^{\prime}(x_{K} - c_{K})]$, which is defined as the distance of the 
$x_{K}$ vector (values of the parameters) from the centre $c_{K}$ where p is the order of the $x_K$ vector. If $d_{K}^{\bf\ \prime}$ is the estimate of $d_{K}$ at the $K^{th}$
point, then the optimum number of clusters is determined by the sharp jump in the curve $J_{K} = (d_{K}^{\bf\ \prime -{p/2}}  - d_{K-1}^{\bf\ \prime-{p/2}})$
vs K \citep{sugar2003}. The jumps as a function of $K$ for our PCA+CA and CA analyses are shown on Fig.~\ref{figjumps}.

\section[]{Analysis with \logs, \logre, \Brie\ and \Mgdeux, and error bars}
\label{fourparam}

%%%%%%%%%%%
   \begin{figure*}
   \centering
 \includegraphics[]{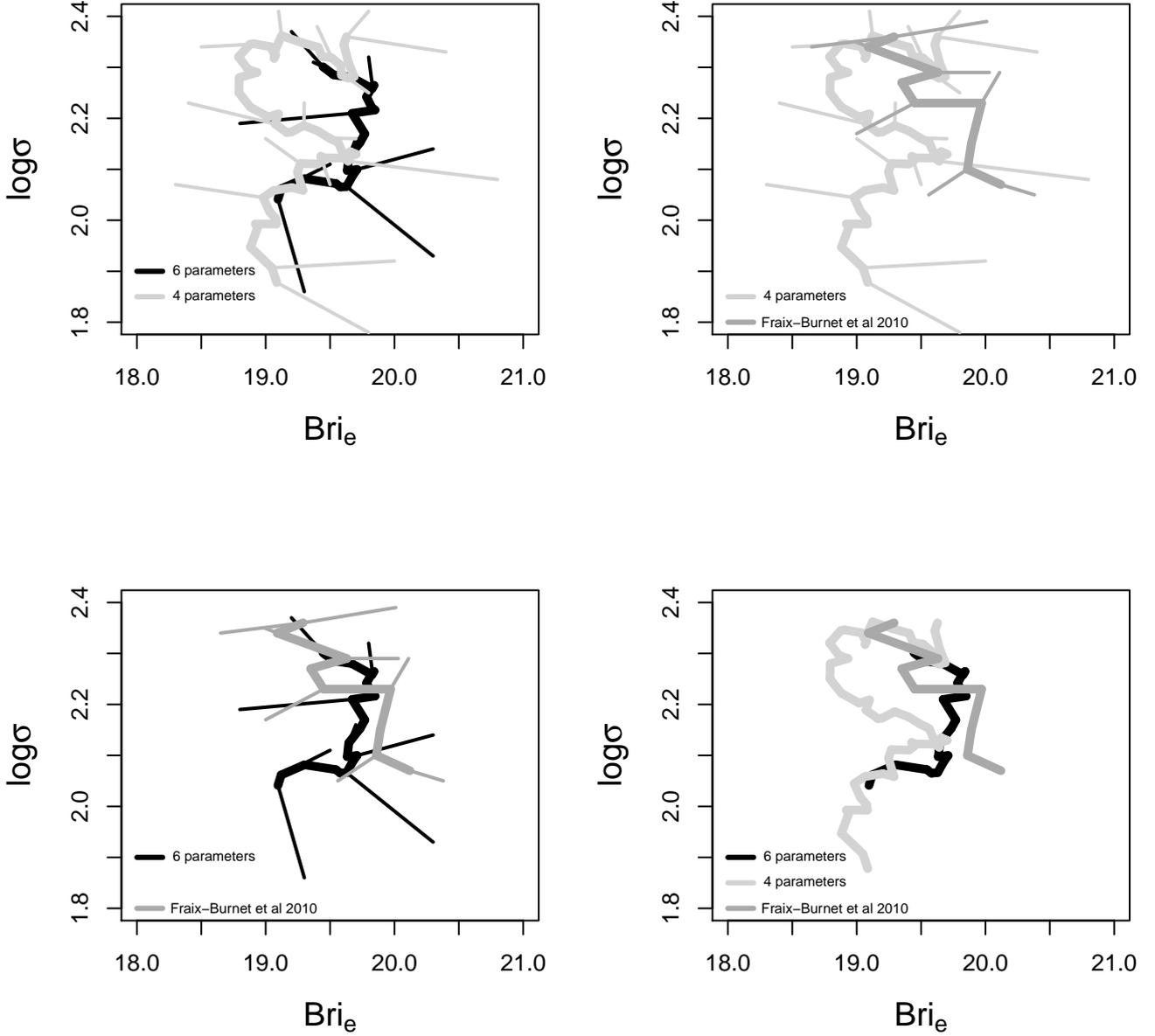}
   \caption{Projection of the trees onto the fundamental plane for three cases: the analysis of this Appendix~\ref{fourparam} and the one by \citet{Fraix2010} both using the four parameters of the fundamental plane, and the principal study of the present paper with six parameters. Thick lines represent the ``trunk'' of the trees, while the small branches relate the trunks to the mean of each group. For clarity, results are compared two by two, and only the trunks are shown for the three studies on the lower right diagram. These are evolutionary tracks in the sense of diversification, and not the path of evolution for a single galaxy.} 
    \label{fig:evoltracks}%
    \end{figure*}
%%%%%%%%%%%%%

\subsection[]{Analysis with \logs, \logre, \Brie\ and \Mgdeux}

We complemented the study presented in this paper with the analysis of our sample with the four parameters (\logre, \logs, \Brie, and $Mg_2$)  as in \citet{Fraix2010}. We used the same three multivariate techniques (cluster analysis, Miminimum Contradiction Analysis, and cladistics) as presented in Sect.~\ref{methods} and Appendix~\ref{appendMeth}.

The resulting tree is less structured (more galaxies lie on individual branches) than the one obtained in the present paper using six parameters. This can be explained by \logre\ and $Mg_2$ not having been found to be discriminant parameters for the considered sample. It is also less structured than in \citet{Fraix2010} which uses the same four parameters, which is probably due to the problems in determining of \logre.

To summarize the results, we show the projection of the three trees -- the one obtained in this paper with six parameters, the one obtained here with four parameters, and the one of \citet{Fraix2010} -- onto the fundamental plane (\logs\ vs \Brie) without the data points (Fig.~\ref{fig:evoltracks}). Globally, there is good agreement and the groupings are consistent. However, the projected tree from the present Appendix departs from the other two in the top half of the figure. This is because this tree is less structured than the others, so that instead of having one or two groups at this level, there is a sequence of single branches that makes the trunk of the tree to ``follow'' more closely individual objects. 

\subsection[]{Influence of $r_e$ and error bars on the partitioning}
\label{errors}

%%%%%%%%%%%
   \begin{figure}
   \centering
 \includegraphics[width=\columnwidth]{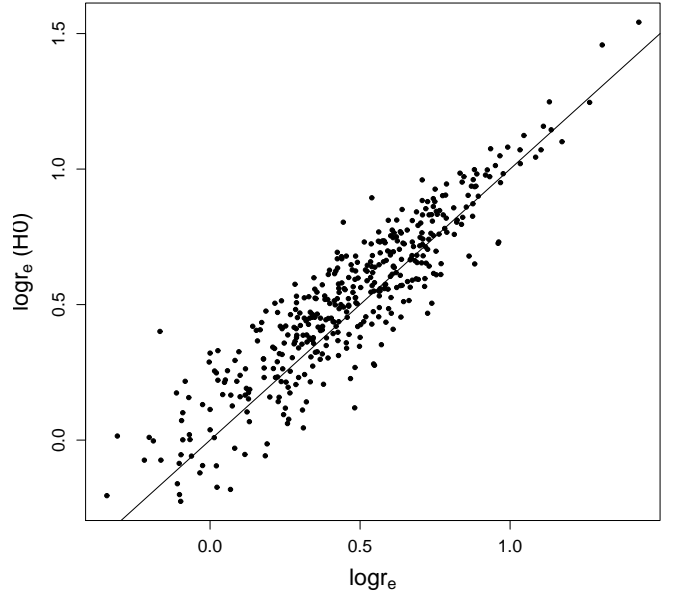}
   \caption{Correspondence between the effective radius computed in two separate ways.} 
    \label{fig:radii}%
    \end{figure}
%%%%%%%%%%%%%

%%%%%%%%%%%
   \begin{figure}
   \centering
 \includegraphics[width=\columnwidth]{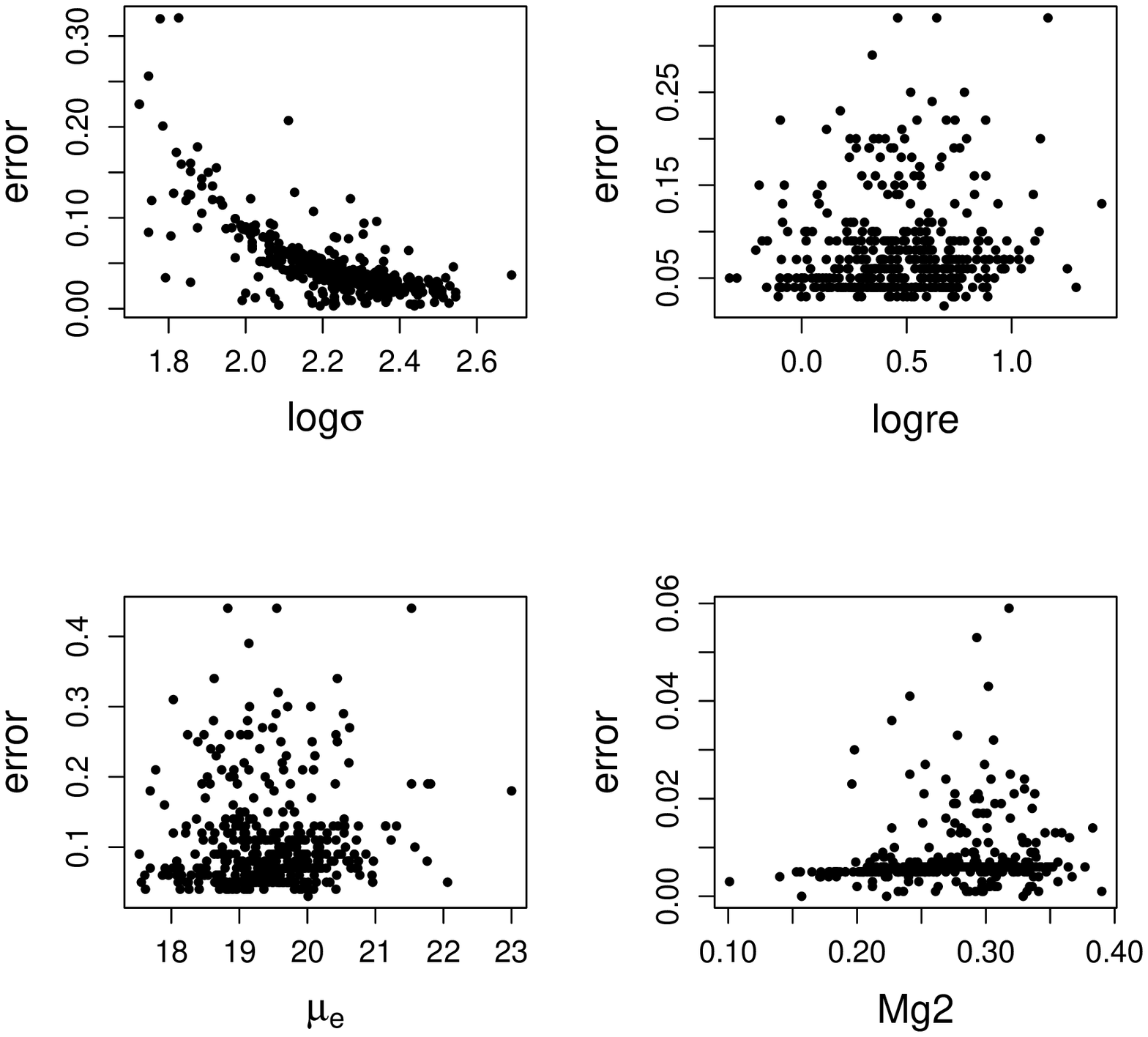}
   \caption{Errors in \logs, \Brie, and \logre\ (taken for the errors in \DsB).} 
    \label{fig:errors}%
    \end{figure}
%%%%%%%%%%%%%

%%%%%%%%%%%
   \begin{figure}
   \centering
 \includegraphics[width=\columnwidth]{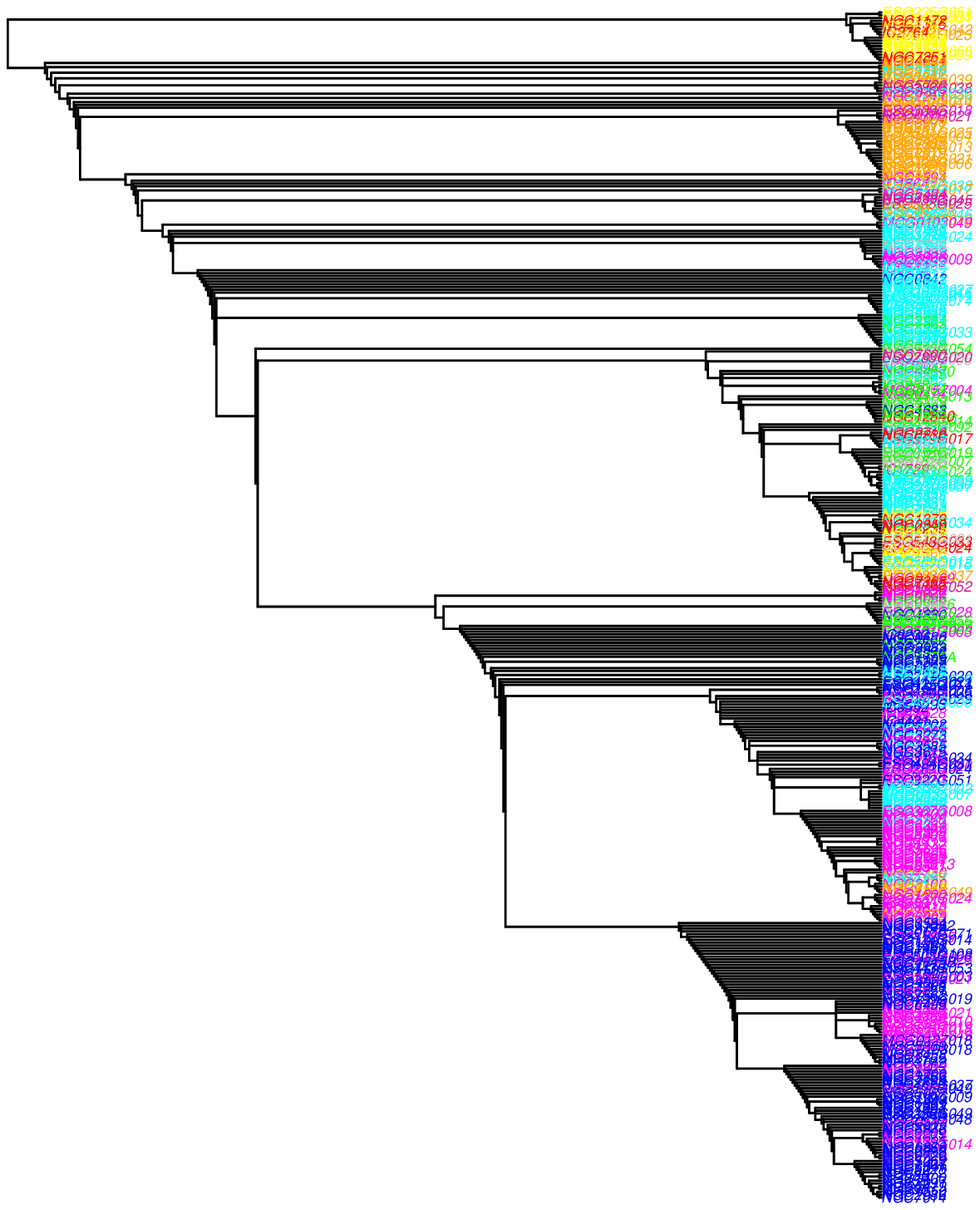}
   \caption{The most parsimonious tree found with cladistics taking uncertainties in the parameters into account. The colours correspond to the groups defined in Fig.~\ref{figtreeclad}.} 
    \label{fig:treesuncertainties}%
    \end{figure}
%%%%%%%%%%%%%

The effective radius \logre\ in our sample is recomputed through a statistical relation between the linear diameter of the galaxy ($D_n$) and its velocity dispersion ($\sigma$), which was determined in another paper \citep{Bernardi2002}. The reason given by \citet{Ogando2008} is that, due to the very low redshift of the galaxies in the sample, ``\textit{the conversion of re in arcseconds to kpc needs a reliable determination of the galaxy distance ($D$). Considering just the redshift to calculate $D$, we may incur in error due to the peculiar motion of galaxies. Thus, we adopted $D$ given by the $D_n$ vs $\sigma$ relation \citep{Bernardi2002} to calculate $r_e$ in kpc.''} However, this relation was obtained with some assumptions (such as the identical properties of galaxies in several clusters) and introduces a dependence of \logre\ (through $D$) on \logs.

The two radii (Fig.~\ref{fig:radii}) are quite well-correlated with each other, but the dispersion is relatively large. We performed two cladistic analyses with the four parameters of the fundamental plane  (\logre, \logs, \Brie, and $Mg_2$) as above using the two determinations of the effective radius. The agreement between the two results is only fair. This can be explained by the relatively important discrepancy between the two different values of $r_e$ (median difference of 10\%). This however is similar to the uncertainty in \logre, but much larger thn for the other parameters. In addition, the radius or dimension of galaxies does not appear as a discriminant parameter in the study presented in this paper. Hence, it is not so surprising that analyses using this parameter are not very stable.

We now consider the robustness of our clustering result for the six-parameter analysis when taking error measurements into account. It is statistically a very challenging task to assess the influence of the errors. However, cladistics can easily take into account the error bars since the optimisation criterion in all analyses performed so far in astrocladistics use the parsimony criterion: among all the possible arrangements of the objects on trees, the simplest evolutionary scenario is retained. The parcimony is measured by using the number of ``steps'', that is the total number of changes in parameter values along all the branches of the tree. If a missing value or an uncertain one (given by a range of values) is included in the data matrix, all possible values are considered and the ones corresponding to the simplest tree is favored. This simply increases the number of possible cases to consider. We note that all possible values within the range alllowed by measurement uncertainties are given the same 
weight, whereas the probability distribution is generally expected to be higher at the central value (ideally gaussian). 

We performed a cladistic analysis similar to that in Fig.~\ref{figtreeclad} using the error bars given in \citet{Ogando2008} and \citet{Alonso2003} for \logs\ and \Brie, and for \DsB\ we considered the error given for \logre\ in \citet{Alonso2003}, There errors are shown in Fig.~\ref{fig:errors}. For \NaD, \MgbFe, and \OIII, we assumed a face value of 10\%, which is the upper limit estimated by \citet{Ogando2008} for all the Lick index values.

The resulting tree shown in Fig.~\ref{fig:treesuncertainties} is slightly less structured than the one in Fig.~\ref{figtreeclad} but most groups are grossly preserved. Clad3 appears to be mixed with Clad1 and Clad5 to be mixed with Clad6. In addition, Clad7 and Clad8 are somewhat mixed with each other. Interestingly, these behaviours are similar to those inferred from the comparison with the partitioning derived from the cluster analysis.
In addition, the agreement is quite satisfactory given the large uncertainties for half of the parameters (the Lick indices), a face value given to these uncertainties, and the equal probability given to all values within the range of uncertainty. 

These results shows that the cladistic analysis is relatively robust to measurement errors, as found through the comparison with different clustering methods.

\section[]{Supplementary figures}
\label{appendFigures}

\clearpage

%%%%%%%%%%%
   \begin{figure}
   \centering
 \includegraphics[width=\columnwidth]{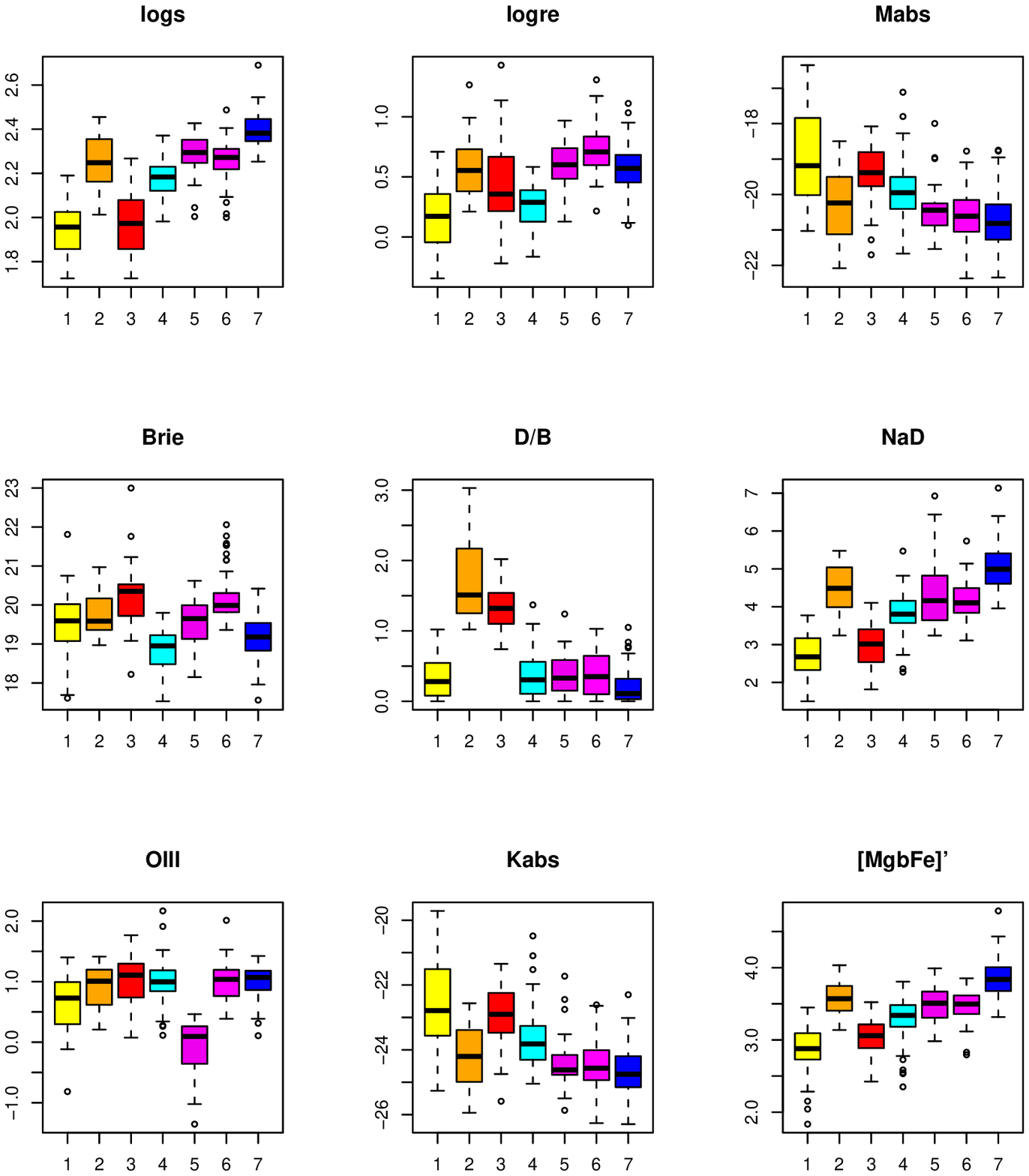}
\includegraphics[width=\columnwidth]{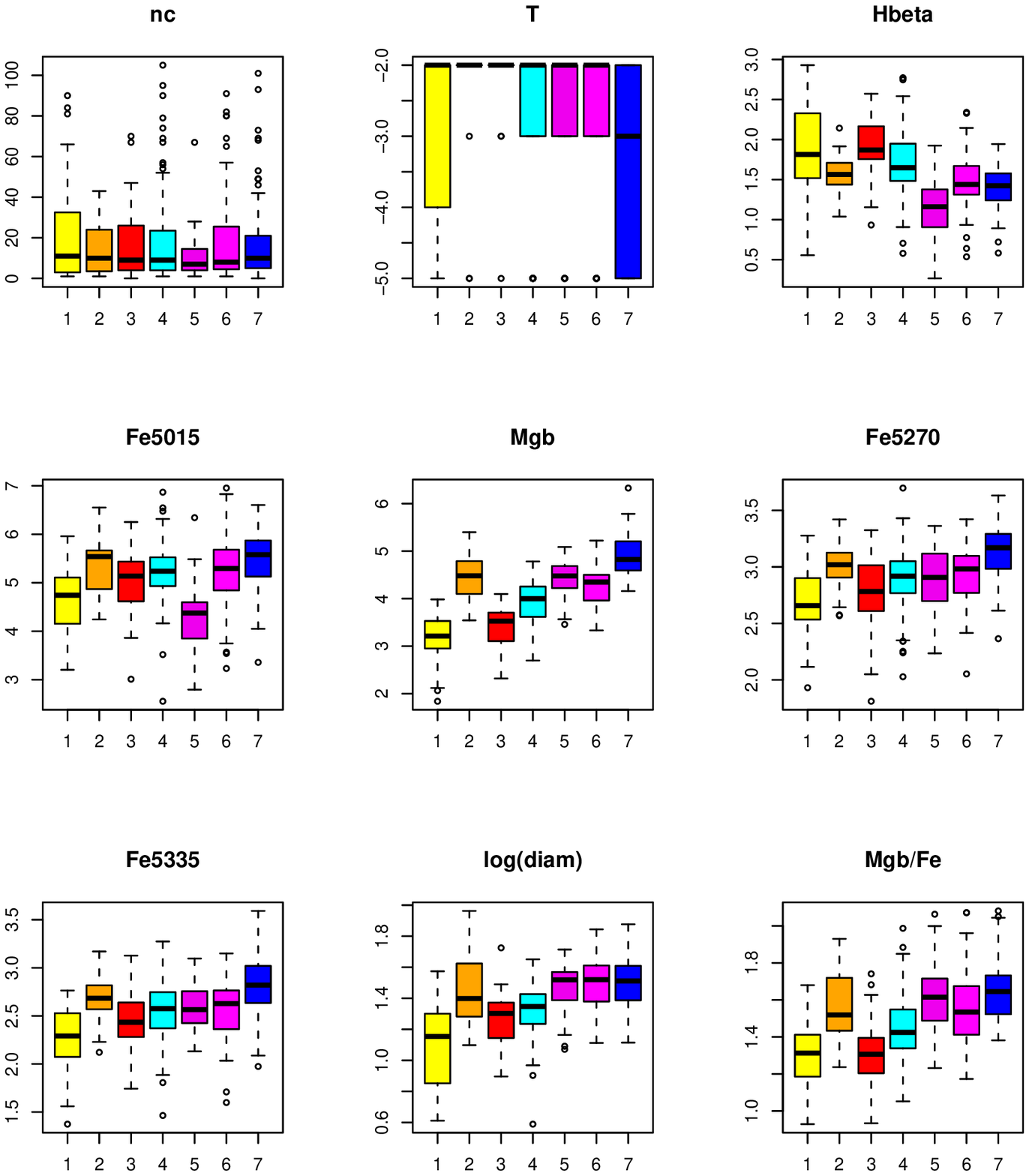} 
 \includegraphics[width=4 true cm]{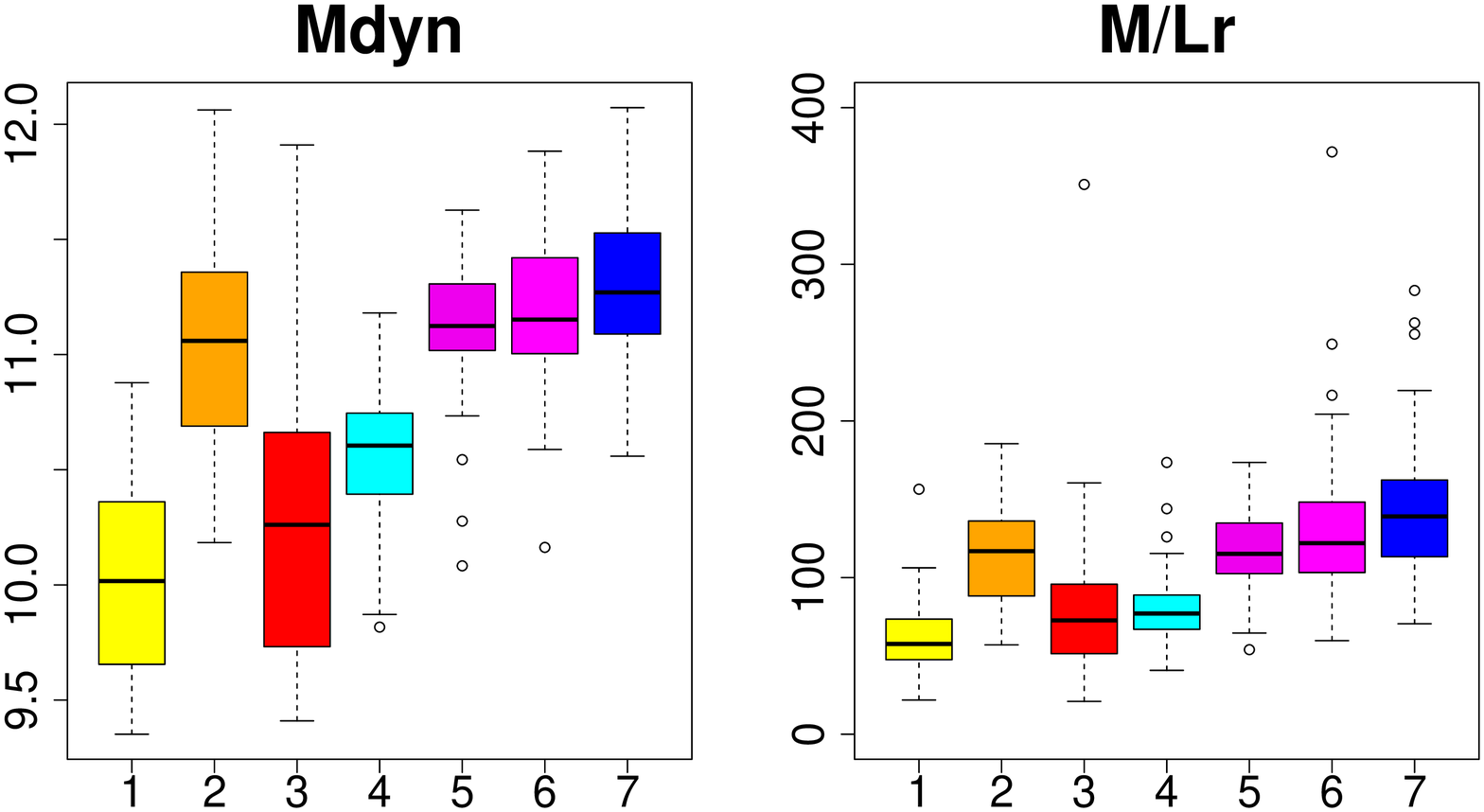}
   \caption{Same boxplots as in Fig.~\ref{figboxplotClad} but for the cluster partitioning. Colours are the one given in Fig.~\ref{figcomppart}.} 
    \label{figboxplotClus}%
    \end{figure}
%%%%%%%%%%%%%
%%%%%%%%%%%
   \begin{figure}
   \centering
 \includegraphics[width=\columnwidth]{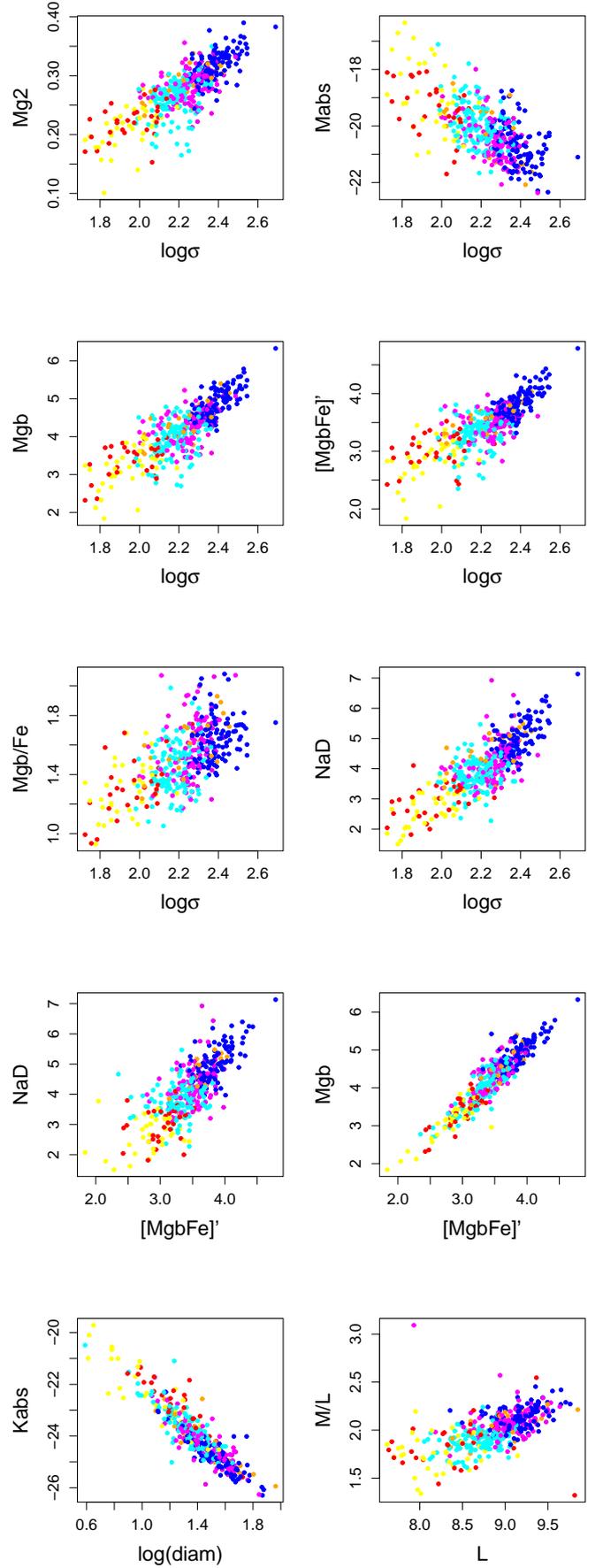}
   \caption{Scatter plots showing evolutionary correlations, like Fig.~\ref{figspurious}, but for the cluster partitioning. Colours are the same as in Fig.~\ref{figboxplotClus}.} 
    \label{figspuriousclus}%
    \end{figure}
%%%%%%%%%%%%%

%%%%%%%%%%%
   \begin{figure*}
   \centering
 \includegraphics[width=16 true cm]{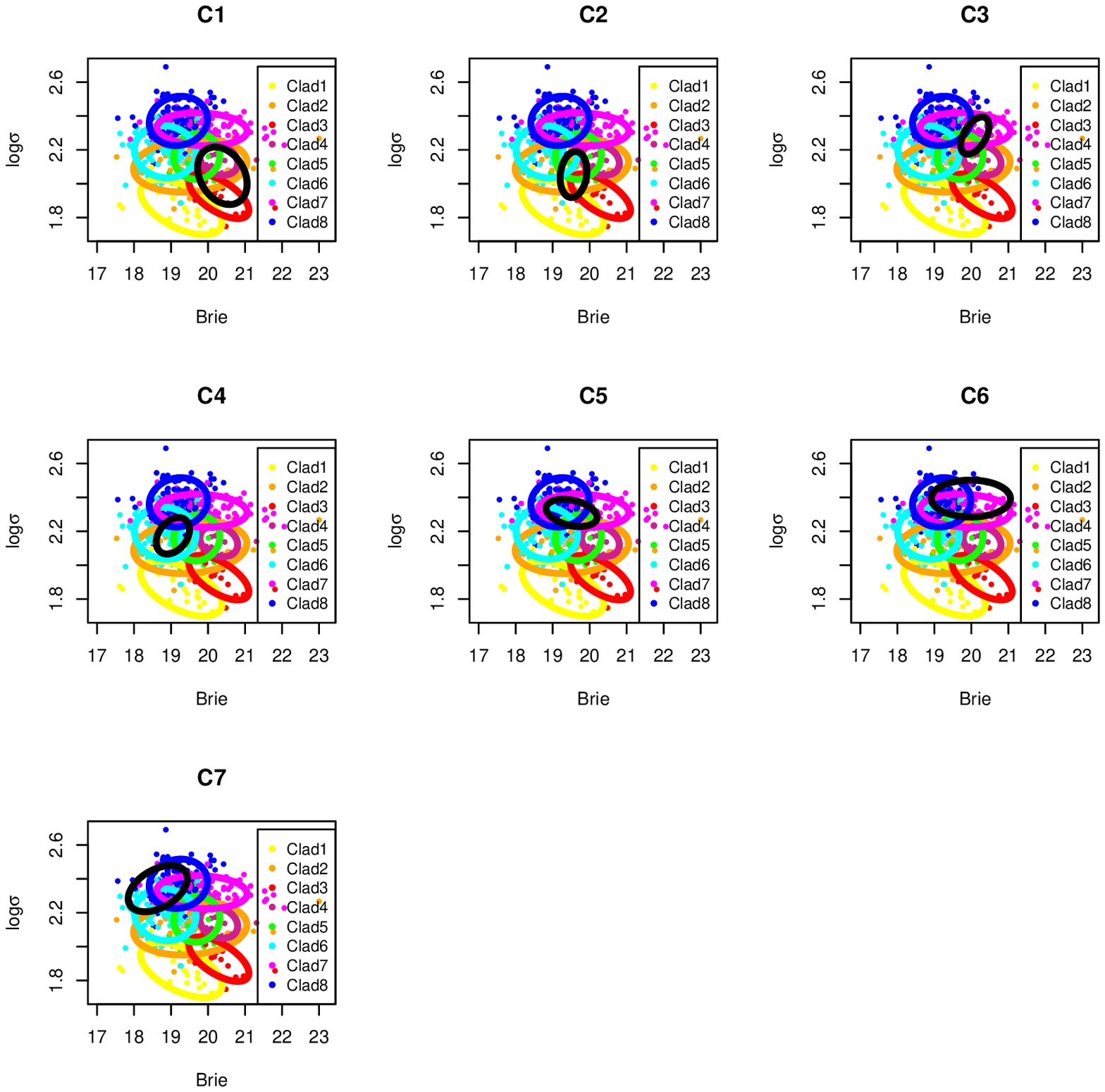}
   \caption{Comparison of the positions of the groups found in \citet{Fraix2010} and those of the present paper, as projected onto the fundamental plane. The colour-coded ellipses are the inertia ellipses for each group from the present paper, and the black ellipse is the one for the group from  \citet{Fraix2010} indicated on top each graph. See also Fig.\ref{figfundplane}.} 
    \label{figcompgroupsFP}%
    \end{figure*}
%%%%%%%%%%%%%
\clearpage
%%%%%%%%%%%
   \begin{figure*}
   \centering
 \includegraphics[width=16 true cm]{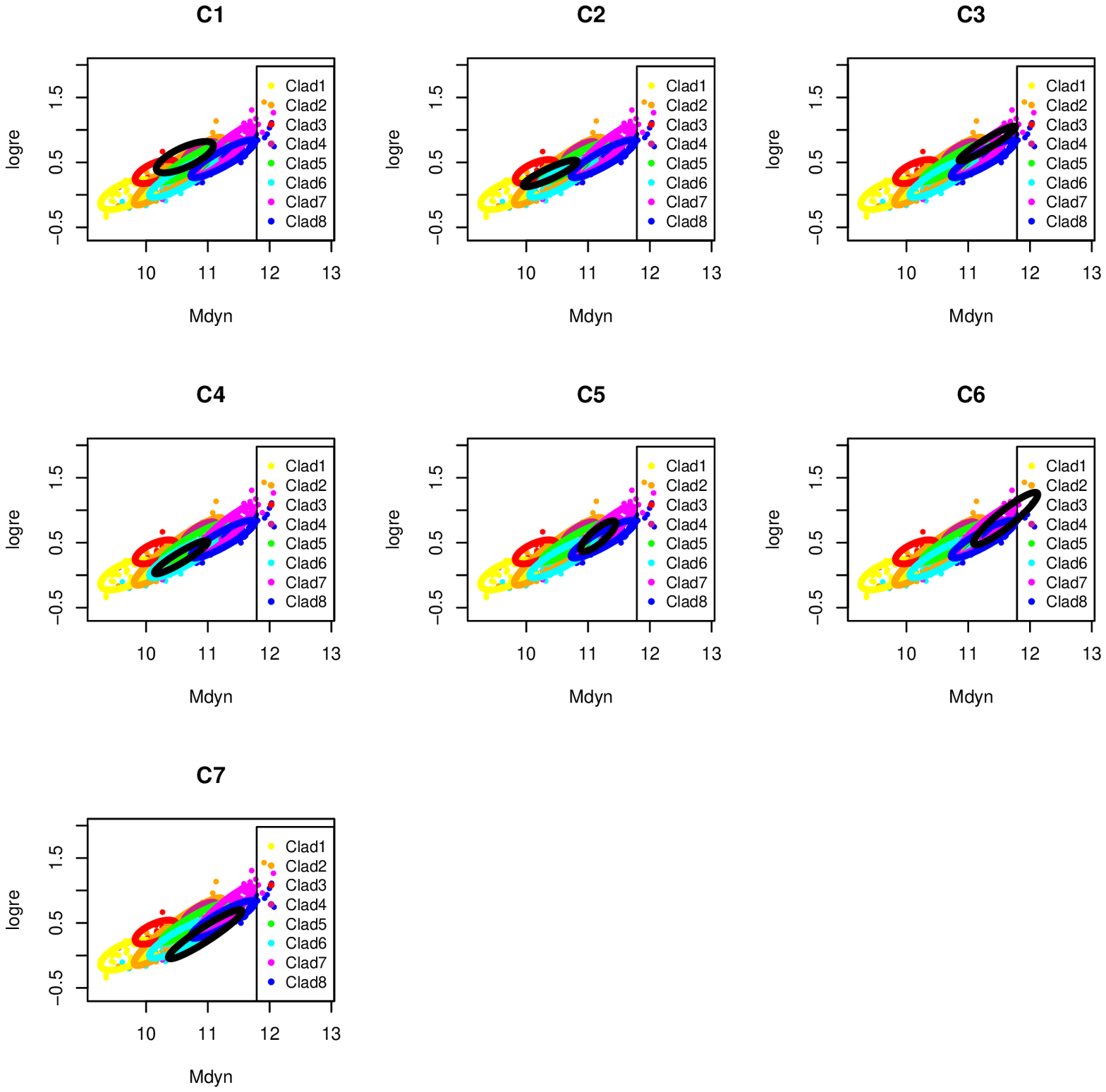}
   \caption{Comparison of the positions of the groups found in \citet{Fraix2010} and those of the present paper, as projected onto the \logre\ vs $M_{dyn}$ diagram. The colour-coded ellipses are the inertia ellipses for each group from the present paper, and the black ellipse is the one for the group from  \citet{Fraix2010} indicated on top each graph. See also Fig.\ref{figcompMdynRe}.} 
    \label{figcompgroupslogreMdyn}%
    \end{figure*}
%%%%%%%%%%%%%

\end{document}